\documentclass[a4paper,fleqn,usenatbib]{mnras}
\usepackage[T1]{fontenc}
\usepackage{ae,aecompl}
\usepackage{multirow}
\usepackage{graphicx}	
\usepackage{amsmath}	
\usepackage{amssymb}	

\usepackage{color}

\newcommand{\be}{\begin{equation}}
\newcommand{\ee}{\end{equation}}
\def\ergs{{\rm\,erg\,s^{-1}}}
\newcommand{\msun}{{M}_{\sun}}
\def\ergcms{{\rm erg cm^{-2}s^{-1}}}

\title[The X-ray spectral evolution of the outbursts in XRBs]{Evolution of hard X-ray photon index in black-hole X-ray binaries: hints for accretion physics}

\author[Liu et al.]{
Hao Liu$^{1}$, AiJun Dong$^{2}$, ShanShan Weng$^{3}$, and Qingwen Wu$^{1}$
\thanks{E-mail: qwwu@hust.edu.cn}
\\
$^{1}$School of Physics, Huazhong University of Science and Technology, Wuhan 430074, China\\
$^{2}$Guizhou Provincial Key Laboratory of Radio Astronomy and Data Processing, Guizhou Normal University, Guiyang 550001, China \\
$^{3}$Department of Physics and Institute of Theoretical Physics, Nanjing Normal University, Nanjing 210023, China\\
}

\date{Accepted XXX. Received YYY; in original form ZZZ}

\pubyear{2016}

\begin{document}
\label{firstpage}
\pagerange{\pageref{firstpage}--\pageref{lastpage}}
\maketitle

\begin{abstract}
The anti- and positive correlations between X-ray photon index and Eddington-scaled X-ray luminosity were found in decay phase of X-ray binary outbursts and a sample of active galactic nuclei in former works. We further systematically investigate the evolution of X-ray spectral index, along the X-ray flux and Eddington ratio in eight outbursts of four black-hole X-ray binaries, where all selected outbursts have observational data from \emph{Rossi X-ray Timing Explorer} in both rise and decay phases. In the initial rise phase, the X-ray spectral index is anti-correlated with the flux, and the X-ray spectrum quickly softens when the X-ray flux is approaching the peak value. In the decay phase, the X-ray photon index and the flux follow two different positive correlations, and they become anti-correlated again when the X-ray flux is below a critical value, where the anti-correlation part follow the same trend as that found in the initial rise phase. Compared with other X-ray binaries, GRO J1655-40 has an evident lower critical Eddington ratio for the anti- and positive transition, which suggests that its black-hole mass and distance are not well constrained or its intrinsic physic is different.  
\end{abstract}


\section{Introduction}
A black hole X-ray binary (XRB) is a gravitationally bound system in which one of the objects is a stellar-mass black hole (BH) with mass from several to $\sim$ twenty solar mass. Apart from several persistent BH XRBs, most of other XRBs are transient sources that can be detected only during outburst. During the outbursts, their luminosities increase several orders of magnitude to reach values close to the Eddington limit ($L_{\rm Edd}=1.38\times10^{38}M_{\rm BH}/\msun$, $M_{\rm BH}$ is BH mass).  At initial stage of the outburst, the XRBs transit from quiescent state to low-hard (LH) state, where the X-ray spectrum is dominated by a power-law component with a photon index of  $\Gamma\sim 1.5-2.1$ and the variability is normally strong accompanying with low-frequency quasi-periodic oscillations (LFQPOs, e.g., $\sim0.1-10$Hz). The X-ray emission was suggested to come from advection-dominated accretion flow \citep[ ADAF,][]{ab95,ny95,esin97,liu07,liu11,yn14}. As the luminosity increases, the XRBs enter into very-high (VH) state (or steep-power-law state, or intermediate state), where the X-ray spectrum becomes very steep ($\Gamma >2.4$), and high-frequency QPO may be appear. The physical origin of the X-ray emission in this state is still unclear, which may correspond to the disk transition. After this stage, the system enters into a high-soft (HS) state, where X-ray spectrum is dominated by a disk component with a weak power-law component ($\Gamma\sim 2.1-2.4$)  and QPO is normally absent. The disk and power-law components come from optically thick standard accretion disk \citep[SSD,][]{shak73}, and the optically thin corona above and below the SSD respectively. Later on, the BH systems return to LH state and, eventually, go back to the quiescent state \citep[e.g., see][for reviews]{remi06,done07,fend12,zhang13}.

The hardness-intensity diagram (HID) is an important tool in understanding the nature of the transient sources \citep[e.g.,][]{remi06,stei16,dunn10,cori11}. The HID evolution mainly reflects the evolution of accretion disk, where the SSD component and the power-law component strongly evolved during one outburst. In exploring the similar physics in different-scale BHs and the corona physics, the X-ray photon index is also widely adopted, which mainly reflect the properties of hot plasma (corona/ADAD/jet) and not suffer the possible contribution from the cold disk. \citet{wu08} performed a spectral study for XRBs in the decay of their outbursts and found that a transition between anti- and positive correlations of  $\Gamma-L_{\rm X}/L_{\rm Edd}$. The anti- and positive correlations between X-ray photon index and Eddington ratio are also found in low-luminosity active galactic nuclei (LLAGNs) and luminous AGNs respectively, where both a single AGN and AGN sample follow these trends \citep[e.g.,][]{wang04,sh06,mar09,sob09,gu09,em12,tr13,al15}. It should be noted that the anti- and positive correlations are roughly consistent with the prediction of ADAF and disk-corona model respectively \citep[e.g.,][]{yu07,cao09,you12,qiao13}.  This roughly scale-free $\Gamma-L_{\rm X}/L_{\rm Edd}$ correlation provides an independent method to estimate the BH mass \citep[e.g.,][]{shap09,jang14} .

The evolution of the X-ray photon index in the rise phase is not well explored in former works, which is partly caused by the quite few observational data in the rise phase, since that the XRB outburst normally follow a fast rise and a slow decay. In this work, we explore the full X-ray spectral evolution for these several XRBs in both rise and decay phase, which will shed light on the evolution of accretion processes and also the evolution of supermassive BH systems (e.g., galaxies). In section 2, we present the description on the sample and the data reduction. The main results are shown in Section 3. In Sections 4 and 5, the discussion and conclusion are presented.   

\section{Sample and Data Reduction}

   We search the \textit{RXTE} database to find the XRB outbursts with the data in both rise and decay phases. Using this criteria, four outbursts of GX 339-4 are selected, which are observed in 2002-2003, 2004-2005, 2006-2007 and 2010-2011, respectively. Two outbursts of H1743-322, that observed in 2003 and 2009, are chosen. Two other outbursts of XTE J1550-564 and GRO J1655-40 that observed in 2000 and 2005 are also included (see Table 1). In total, eight outbursts of four BH XRBs are chosen, where the distance, BH mass and other related information are presented in Table 1.
   
\begin{table*}
\caption{X-ray binary parameters\label{tab:pks142p220}.}
\begin{tabular}{cccccc}
\hline \hline
  Object & Outburst & Observation & BH mass & Distance & $N_{\rm H}$$\times$$10^{22}$   \\
   name &  year   & Number  & ($\msun$) & (kpc) & (cm$^{-2}$)   \\
  \hline
  GX 339-4 & 2002-03 & 151 & $ 9.0 _{ -1.2 }^{+ 1.6 }$(1) & $8.4\pm0.9$(1) & 0.5(2)   \\
    & 2004-05 & 97 & & &  \\
    & 2006-07 & 190 & & &  \\
    & 2010-11 & 158 & & &  \\
    \hline
  XTE J1550-564 & 2000 & 67 & 10.0(3) & 5.0(4) & 2.0(5)  \\
  \hline
  H1743-322 & 2003 & 151 & 10.0(6) & 11.0(7) & $1.8\pm0.2$(8) \\
    & 2009 & 50 & & &  \\
    \hline
  GRO J1655-40 & 2005 & 165 & $6.3\pm0.5$(9) & $3.2\pm0.2$(10) & 0.8(11) \\

\hline \hline
\end{tabular}
\begin{minipage}{120mm}
References: (1) \cite{park16}; (2) \cite{mott09}; (3) \cite{oros02}; (4) \cite{kale02}; (5) \cite{tita02}; (6) \cite{mill12}; (7) \cite{kale06}; (8) \cite{cori11}; (9) \cite{gree01}; (10) \cite{hjel95}; (11) \cite{kale16}.
\end{minipage}

\end{table*}

The X-ray data is reduced and analyzed using HEASOFT software (version 6.25) following the steps as described in RXTE cookbook\footnote{http://heasarc.gsfc.nasa.gov/docs/xte/data-analysis.html}. In order to uniformly explore the X-ray spectral evolution,  we consider the \emph{Proportional Counter Array} (PCA) data and the \emph{High-Energy X-ray Timing Experiment} (HEXTE) data in 3-25keV and 20-200keV, respectively.  For PCA data, we use only the data from PCU-2 as it is always switched on, and can be used over the entire archive of data, which is also the best calibrated of the PCUs on RXTE.  Because the Clusters A and B in HEXTE are ceased modulation in 2006 and 2009 respectively \citep[see][for details]{mott11}, we use the HEXTE data from Cluster B for seven of eight outbursts in our sample. For outburst 2010-2011 of GX 339-4, only PCA data are adopted due to no HEXTE observations. We extract the useful observing time periods, known as good time intervals (GTIs), using the ftool MAKETIME with the following observational constraints: elevation angle is larger than 10 degree, and spacecraft pointing offset is less than 0.02. Response matrices are generated and background spectra are created using the latest PCA background model (faint or bright) according to brightness level.

The spectra are fitted in XSPEC (version 12.10.1). The X-ray spectrum can roughly be modeled by a power-law component and a possibly blackbody component in different states. As our spectral analysis concentrates on the X-ray spectral evolution, we make the model as simple as possible. Therefore, we adopt a power-law component (POW) and an absorption component (PHABS) as a starting model, a disk component (DISKBB), a Gaussian line fixed at 6.4 keV and a high energy cutoff component (CUTOFF) will be added if they can improve the fittings substantially. As shown in Table 1,  the hydrogen column density was fixed at $N_{\rm H}=0.5\times10^{22}\rm cm^{-2}$ for GX 339-4 \citep[e.g.,][]{mott09}, $N_{\rm H}=0.8\times10^{22}\rm cm^{-2}$ for GRO J1655-40 \citep[e.g.,][]{kale16}, $N_{\rm H}=1.8\times10^{22} \rm cm^{-2}$ for H1743-322 \citep[e.g.,][]{cori11}, and $N_{\rm H}=2.0\times10^{22}\rm cm^{-2}$ for XTE J1550-564 \citep[e.g.,][]{tita02}. A systematic uncertainty of 0.5\% to all channels is adopted in our work to account for PCA and HEXTE calibration uncertainties, where we get a satisfactory fit for a selected model.  In some HS states, the power-law component is very weak, and, therefore, the X-ray photon index cannot be well constrained. In this work, we focus on the evolution of the X-ray photon index with the unabsorbed X-ray flux, and, therefore, we neglect the data points that the flux of power-law component is less than 10\% of total flux in some HS states. The observational IDs, the observational date, main fitting parameters of disk and power-law components, 3-25keV X-ray flux and adopted models for each outbursts are shown in Tables 3--10 respectively.

\section{Results}

  We present the light curve in 3-25 keV waveband and the relation of $\Gamma-F_{\rm 3-25 keV}$ for eight outbursts of four sources in the left and right panels of Figures 1--8 respectively. The grey points represent the observations with quite weak power-law component (less than 10\% of total flux). The X-ray spectral evolutions are more or less similar for different outbursts, even the observational data may be missed in initial stage of some outbursts. We summarize the main features of the X-ray spectral evolution for most outbursts, where more details are also provided in Table 2. In the initial rise phase of the outbursts, $\Gamma$ and $F_{\rm 3-25 keV}$ follow an anti-correlation when the X-ray flux is less than a critical flux (e.g., dark red points, see Figures 1--8), where the anti-correlation can be extended to very bright hard state with $\Gamma\sim 1.3$ when the flux is approaching the peak value. Then, the X-ray spectrum quickly softens with roughly unchanged X-ray flux (e.g., $\Gamma\sim 1.3 - 2.5$). After the peak X-ray flux, the sources enter into the decay phase. The X-ray photon index and X-ray flux follow a  shallow positive correlation, where the sources mainly stay in the HS state. With further decrease of the X-ray flux, X-ray spectrum show a second strong variation, where the X-ray photon index varies from $\Gamma\sim 2.5$  to 1.5 even the X-ray flux only changes a little bit. At the end of decay phase, $\Gamma$ and $F_{\rm 3-25 keV}$ follow an anti-correlation again, where the anti-correlations roughly follow the same trend as that found in the initial rise phase. It should be noted that the X-ray spectrum in the rise phase can be extended to a harder spectrum (e.g., $\Gamma_{\rm min}\sim 1.3$) at higher critical flux compared that in the decay phase ($\Gamma_{\rm min}\sim 1.6$). It should be cautious for the X-ray spectral evolution of outburst 2010-2011 of GX 339-4 (Figure 4), where HEXTE data are lacked and the photon index may be a little bit different from other outbursts with HEXTE observations (e.g., cutoff power-law model is adopted in some LH states with HEXTE data). Some outbursts lack the observation in the initial rise phase, and the anti-correlations of $\Gamma-F_{\rm 3-25 keV}$ are absent (e.g., outbursts of 2004-2005, 2010-2011 for GX 339-4, outburst 2000 of XTE J1550-564, outbursts 2003 and 2009 of H1743-322).  
  
  To learn the X-ray spectral evolution for these outbursts as a whole, we present the relation of $\Gamma-L_{\rm 3-25 keV}/L_{\rm Edd}$ for these eight outbursts in Figure 9. We find that four outbursts of GX 339-4 and XTE J1550-564 roughly follow the same track evolution, except that the critical fluxes for steep positive correlations of $\Gamma-L_{\rm 3-25 keV}/L_{\rm Edd}$ in the rise phase are different for different outbursts. We find the X-ray photon index from the anti-correlation part of $\Gamma-L_{\rm 3-25 keV}/L_{\rm Edd}$ in H1743-322 is evidently softer than those of other outbursts at given Eddington ratio.  The anti-correlation of $\Gamma-L_{\rm 3-25 keV}/L_{\rm Edd}$ in GRO J1655-40 is quite consistent with other outbursts (e.g., GX 339-4 and XTE J1550-564), and, however, the positive correlations of $\Gamma-L_{\rm 3-25 keV}/L_{\rm Edd}$ in both rise phase and decay phase are evidently different from others, where the critical Eddington ratio for the transition of anti- and positive correlations is about one order of magnitude lower than those of others. 

\begin{table*}
\caption{The summary of the $\Gamma$ evolution  for the different outbursts of XRBs.\label{tab:pks142p220}}
\begin{tabular}{cccccc}
\hline \hline
  Outburst &  &  & $F_{\rm 3-25keV}$ & $\Gamma$  & Correlation  \\
   year  & Phase & Date(MJD) & $\ergcms$ &  &  \\
  \hline
   \multicolumn{6}{c}{\textbf{GX 339-4}} \\
   2002-03 & rise & 52324.39--52367.76 & $<3.5\times10^{-9}$ & $\sim$ 2.1 to $\sim$ 1.1 & negative  \\
      &  & 52367.76--52412.06 & $3.5\times10^{-9}$ to $\sim1.3\times10^{-8}$ & $\sim$ 1.1 to $\sim$ 2.5 & positive  \\
      & decay & 52412.06--52686.25 & $1.2\times10^{-8}$ to $1.0\times10^{-9}$ & $\sim$ 2.5 to $\sim$ 2.3  & positive  \\
      &  & 52686.25--52739.58 & $\simeq 1.0\times10^{-9}$ & $\sim$ 2.3 to $\sim$ 1.5  & positive  \\
      &  & 52739.58--52826.06 & $<8.0\times10^{-10}$ & $\sim$ 1.5 to $\sim$ 2.1  & negative  \\ 
     \cline{2-6}
     2004-05 & rise & 53225.40--53234.58 & $\simeq 4.0\times10^{-9}$ & $\sim$ 1.4 to $\sim$ 2.3  & positive  \\
      & decay & 53234.58--53466.74 & $5.0\times10^{-9}$ to $\sim1.0\times10^{-9}$ & $\sim$ 2.3 to $\sim$ 2.2  & positive  \\
      &  & 53466.74--53490.55 & $\simeq 1.0\times10^{-9}$ to $4.0\times10^{-10}$ & $\sim$ 2.2 to $\sim$ 1.6  & positive  \\
      &  & 53490.55--53539.05 & $<4.0\times10^{-10}$ & $\sim$ 1.6 to $\sim$ 2.1  & negative  \\
     \cline{2-6}
     2006-07 & rise & 54051.74--54097.05 & $<3.0\times10^{-9}$ & $\sim$ 1.7 to $\sim$ 1.1  & negative \\
      &  & 54097.05--54136.99 & $3.0\times10^{-9}$ to $1.3\times10^{-8}$ & $\sim$ 1.1 to $\sim$ 1.3  & positive  \\
      &  & 54136.99--54147.01 & $\simeq 1.3\times10^{-8}$ & $\sim$ 1.3 to $\sim$ 2.5  & positive  \\
      & decay & 54147.01--54217.27 & $1.2\times10^{-8}$ to $1.0\times10^{-9}$ & $\sim$ 2.5 to $\sim$ 2.2  & positive  \\
      &  & 54217.27--54244.96 & $\simeq 1.0\times10^{-9}$ & $\sim$ 2.2 to $\sim$ 1.5  & positive  \\
      &  & 54244.96--54299.49 & $<1.0\times10^{-9}$ & $\sim$ 1.5 to $\sim$ 1.7  & negative  \\
     \cline{2-6}
     2010-11 & rise & 55208.48--55290.72 & $1.0\times10^{-9}$ to $1.0\times10^{-8}$ & $\sim$ 1.3 to $\sim$ 1.6  & positive  \\
      &  & 55290.72--55323.21 & $\simeq 1.0\times10^{-8}$ & $\sim$ 1.6 to $\sim$ 2.5  & positive  \\
      & decay & 55323.21--55578.88 & $1.0\times10^{-8}$ to $1.0\times10^{-9}$ & $\sim$ 2.5 to $\sim$ 2.2  & positive  \\
      &  & 55578.88--55609.84 & $\sim 1.0\times10^{-9}$ to $5.0\times10^{-10}$ & $\sim$ 2.2 to $\sim$ 1.6  & positive  \\
      &  & 55609.84--55639.50 & $<5.0\times10^{-10}\ergs$ & $\sim$ 1.6 to $\sim$ 1.8  & negative  \\
    \hline
    \multicolumn{6}{c}{\textbf{XTE J1550-564}} \\
    2000 & rise & 51644.47--51662.16 & $<1.5\times10^{-8}$ & $\sim$ 1.3 to $\sim$ 2.2 & positive  \\
      & decay & 51662.16--51686.29 & $1.5\times10^{-8}$ to $3.0\times10^{-9}$ & $\sim$ 2.2 to $\sim$ 1.5  & positive  \\
      &  & 51686.29--51741.39 & $<3.0\times10^{-9}$ & $\sim$ 1.5 to $\sim$ 2.1  & negative  \\
  \hline
  \multicolumn{6}{c}{\textbf{H1743-322}} \\
    2003 & rise & 52726.80--52786.28 & $\sim 1.9\times10^{-9}$ to $2.8\times10^{-8}$ & $\sim$ 1.1 to $\sim$ 2.6  & positive  \\
      & decay & 52786.28--52807.60 & $2.8\times10^{-8}$ to $1.2\times10^{-8}$ & $\sim$ 2.6 to $\sim$ 2.1  & positive  \\
      &  & 52932.09--52945.17 & $1.5\times10^{-9}$ to $6.0\times10^{-10}$ & $\sim$ 2.1 to $\sim$ 1.8  & positive  \\
      &  & 52945.17--52959.22 & $<6.0\times10^{-10}$ & $\sim$ 1.8 to $\sim$ 2.3  & negative  \\
     \cline{2-6}
     2009 & rise & 54980.39--54989.15 & $\simeq 4.0\times10^{-9}$ & $\sim$ 1.3 to $\sim$ 2.3  & positive  \\
      & decay & 54989.15--55028.88 & $5.0\times10^{-9}$ to $5.0\times10^{-10}$ & $\sim$ 2.3 to $\sim$ 1.9  & positive  \\
      &  & 55028.88--55055.60 & $<5.0\times10^{-10}$ & $\sim$ 1.8 to $\sim$ 2.2  & negative  \\
    \hline
    \multicolumn{6}{c}{\textbf{GRO J1655-40}} \\
    2005 & rise & 53426.04--53433.90 & $<1.2\times10^{-9}$ & $\sim$ 1.5 to $\sim$ 1.3  & negative  \\
      &  & 53433.90--53507.72 & $1.2\times10^{-9}$ to $\sim 8.0\times10^{-8}$ & $\sim$ 1.3 to $\sim$ 2.8  & positive  \\
      & decay & 53507.72--53539.38 & $8.0\times10^{-8}$ to $1.5\times10^{-8}$ & $\sim$ 2.7 to $\sim$ 2.0  & positive  \\
      &  & 53628.91--53636.18 & $\sim 4.0\times10^{-9}$ to $9.0\times10^{-10}$ & $\sim$ 2.0 to $\sim$ 1.5  & positive  \\
      &  & 53636.18--53657.15 & $<9.0\times10^{-10}$ & $\sim$ 1.5 to $\sim$ 2.0  & negative  \\

\hline \hline
\end{tabular}

\end{table*}

\section{Discussion}
The X-ray spectral evolution during the decay phase of XRBs has been well studied, where the anti- and positive correlations of $\Gamma-L_{\rm X}/L_{\rm Edd}$ are found for the Eddington-scaled X-ray luminosity is lower and larger than $\sim1\%$ respectively \citep[e.g.,][]{yu07,wu08}.  We present the evolution of the X-ray photon index for 8 outbursts in both rise and decay phase based on the RXTE observations of 4 XRBs, where the X-ray spectral evolution is more complex compared the results as reported in former works. In the rise phase of the outbursts, $\Gamma$ and $F_{\rm 3-25 keV}$ are anti-correlated at initial stage, where the XRBs change from the quiescent state to bright hard state ($\Gamma$ varies from $\sim$2.0 to $\sim$1.3). Then, the X-ray spectrum quickly softens when the flux is approaching the peak flux ($\Gamma$ varies from $\sim$1.3 to $\sim$2.5), and the XRBs enter into the VH state. In the decay phase, the $\Gamma$ and $F_{\rm 3-25 keV}$ follow a shallower positive correlation as decreases of the flux and the sources transit from the VH to HS state, and the X-ray spectrum harden quickly again when the source change from the HS state to LH state. After a critical flux or Eddington ratio, $\Gamma$ and $F_{\rm 3-25 keV}$ follow an anti-correlation, which is similar to that as found in initial rise phase. Compared the decay phase, the anti-correlation $\Gamma$ and $F_{\rm 3-25 keV}$ can extend to a harder spectrum with a higher critical X-ray flux.

The former works showed that the critical luminosity for the transition from soft state to hard state is roughly constant during the decay phase in BH XRBs \citep[$\sim 2\%L_{\rm Edd}$, e.g.,][]{ma03,gl07} even the critical luminosity is different for different outbursts during the hard to soft transition in the rise phase \citep[e.g., ][]{yu09,yan15}.  With the adopted BH mass and distance of GRO J1655-40, we find that the critical Eddington ratio for the transition of anti- and positive correlation of $\Gamma-L_{\rm 3-25 keV}/L_{\rm Edd}$ in both rise phase and decay phase is several times lower than that of other sources. The lower critical Eddington ratio during the decay phase of GRO J1655-40 is either caused by the uncertainties of basic parameters (e.g., distance and BH mass) or caused by the different intrinsic physical condition in accretion-jet physics.  The BH mass is estimated from several methods (e.g., dynamical method, empirical method from the quasi-periodic oscillation etc.), and the BH mass are restricted within a range of 5--7$\msun$ \citep[e.g.,][]{beer02,oros97,stuc16}, which is not far from our adopted value \citep[e.g.,][]{gree01}. The distance of $3.2\pm1.2$ kpc is normally adopted for GRO J1655-40 in the literatures, which is estimated from a kinematic model of the radio jets \citep[e.g.,][]{hjel95}.  \citet{foe06} determined a spectral type for the secondary star during the quiescence and proposed that the distance should be smaller than 1.7 kpc by comparing this companion with various stars of similar spectral types. If this is the case, the critical value for the transition of anti- and positive correlation should be further 3--4 times lower than our derived value in this work, and will strengthen that this source may be different from other sources. More outburst analysis and better constraints on its basic parameters will help to understand this issue.  In H1743-322, the X-ray spectrum in the anti-correlation part is evidently softer than other sources at given Eddington ratio, which may be caused by observational bias \citep[e.g., Galactic ridge emission due to close to the Galactic plane with $b = 1.8^{\circ}$,][]{kale06,dinc12} or different initial conditions in accreting material (e.g., stronger magnetic field leads to lower electron temperature and softer spectrum). It should be noted that the anti-correlation still exist even considered the Galactic ridge emission \citep[see Figure 2 in][]{kale06}, and this emission will not affect the spectral fittings in bright hard state, high/soft state and very high state. It is also proposed that the X-ray spectrum may be inclination-dependent \citep[e.g.,][]{muno13,hei15}. In Figure 9, we find GX 339-4, XTE J1550-564 and GRO J1655-40 follow the similar anti-correlation of $\Gamma-L_{\rm 3-25 keV}/L_{\rm Edd}$, where their inclination angles range from $60^{\circ}$ to $75^{\circ}$  \citep[e.g.,][]{hei15}. The inclination-dependent X-ray spectrum should be less important in quiescent to low/hard state. The inclination angle of H1743-322 is similar to that of XTE J1550-564, and the difference in the X-ray spectral evolution should be caused by other effects. The combination of timing and X-ray spectral properties may help to explore the physical reasons for different X-ray spectral evolution(e.g., accretion-jet properties, inclination effect etc.).

The anti- and positive correlations of $\Gamma-L_{\rm X}/L_{\rm Edd}$ are found in both XRBs and a sample of AGNs, which are explained by the ADAF and disk-corona model respectively \citep[e.g.,][]{wang04,sh06,yu07,wu08,mar09,sob09,gu09,em12,tr13,plot13,al15,yang15}. The X-ray photon index, $\Gamma$, is regulated by the electron temperature, $T_{\rm e}$, and optical depth, $\tau$, of the corona/ADAF \citep[the so-called y-parameter, e.g.,][]{zd96}. The ADAF model predicts the soften of the X-ray spectrum as XRBs fading into quiescent state (or above anti-correlation), where the optical depth for Comptonization decreases, decreasing the Compton y-parameter, thereby leading to a softer X-ray spectrum \citep[e.g.,][]{qiao13,yang15}. As the accretion rate increases, the hot plasma in ADAF/corona will be cooled down into SSD or cold clumps and the optical depth of corona will decrease quickly, which leads to a smaller y-parameter and a softer X-ray spectrum \citep[e.g.,][]{cao09,you12,qiao13,yang15}. The change of the slope in two positive correlations during the decay phase is unclear, which is also different from that found in AGNs where a single positive correlation is normally reported \citep[e.g.,][]{wang04,sh06}. \citet{cao09} found that the evolution of X-ray photon index is closely correlated to the underlying magnetic stress in the disk-corona model, where the X-ray spectral index will saturate if the magnetic stress is proportional to pure gas pressure(their top panel of Figure 4).  If this is the case, our X-ray spectral evolution can help to constrain the underlying magnetic stress in the accretion disk, which will be our future work.

The physical mechanism behind the hysteresis (e.g., HID or $\Gamma$--Flux diagram) is still not well understood. The candidate explanations include the magnetic field in accretion disk \citep[e.g., ][]{bh08,ba14,cao16} or the instabilities associated with Lense-Thirring effect \citep[e.g., ][]{ns14}. Two outbursts of GX 339-4 at 2002-2003 and 2006-2007 are roughly observed from the quiescent state, where the anti-correlation part is roughly similar to that in decay phase except for the higher critical flux. The higher critical X-ray flux in the rise phase may be caused by the increase of the radiative efficiency in the accretion disk if magnetic field became stronger at this stage, and different initial magnetic field condition in accreting material will lead to different critical X-ray flux \citep[e.g., ][]{ba14,cao16}.  However, the transition from SSD to ADAF in decay phase is caused by much different physical process (e.g., evaporation). The detailed spectral calculations based on above different models are still absent, which will be crucial to understand the evolution of the BH central engine and the hysteresis effect.

\section{Conclusion}

We analyze the full evolution of the X-ray photon index along the X-ray flux or Eddington ratio for eight outbursts of four XRBs, which have the observations in both the rise and decay phases. We find that X-ray spectral evolution is much more complex compared that in the decay phase of XRBs and AGN samples as reported in former works. The main results are summarized as follows: 1) In the decay phase, there are two positive correlations of $\Gamma$-Flux, which is not reported in former works of XRBs and AGNs, and may correspond to the physically changed evolution of the accretion disk; 2) The X-ray emission can be extended to a harder spectrum (e.g., $\Gamma_{\rm min}\sim 1.3$) at a higher critical flux in rise phase compared that in the decay phase ($\Gamma_{\rm min}\sim 1.6$); 3) The anti-correlation of $\Gamma$-Flux follow the same trends in both rise and decay phase, which suggests that their accretion process should be similar in this stage; 4) We find that the critical Eddington ratio of anti- and positive  $\Gamma-F_{\rm 3-25 keV}$ correlation for GRO J1655-40 is evidently lower than that of other sources, which suggests that its BH mass and distance are not well constrained or intrinsic physics is different. More theoretical work need to further study these properties, which can shed light on the evolution of accretion process and the hysteresis effect.

\section*{Acknowledgements}
This work is supported by the NSFC (grants 11622324, 11573009 and 11673013), and Doctor Starting Up Foundation of Guizhou Normal University (0516134).

\begin{figure*}
\centering
\includegraphics[width=17cm,height=6.2cm]{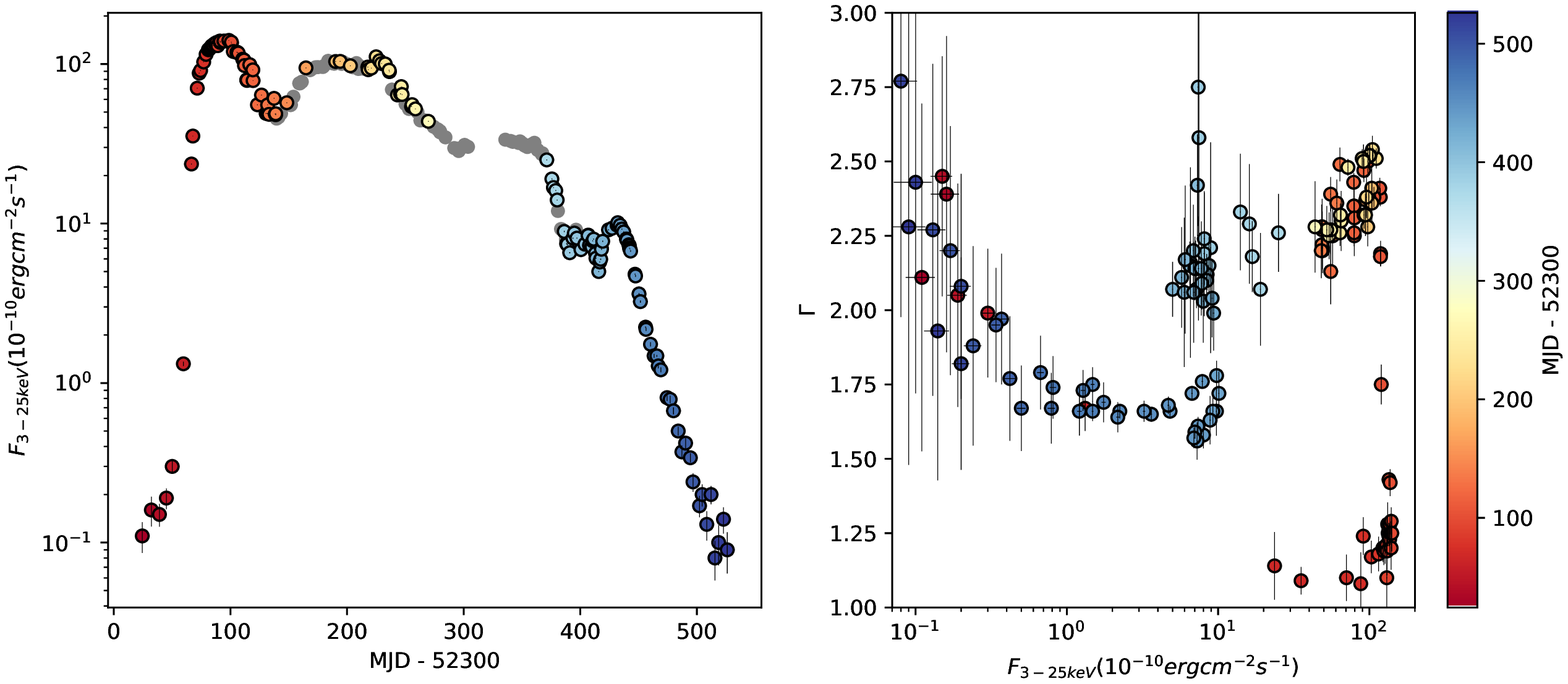}
\caption{Light curve and $\Gamma-F_{\rm 3-25 keV}$ relation for outburst 2002-2003 of GX 339-4. The circle points with different colors represent the data points in different time. The red-like points represent the anti-correlation, shallow and steep positive correlation of $\Gamma-F_{\rm 3-25 keV}$ in the rise phase. The blue-like points represent the shallow positive correlation, steep positive correlation and the anti-correlation of $\Gamma-F_{\rm 3-25 keV}$ in the decay phase. The grey points in left panel represent either the observations with power-law component is less than 10\% of total flux or the observations can not well fitted by our adopted simple model. The 3-25 keV X-ray flux is in unit of $10^{-10}\rm erg cm^{-2} s^{-1}$.
}
\label{fig:2009p2013}
\end{figure*}

\begin{figure*}
\centering
\includegraphics[width=17cm,height=6.2cm]{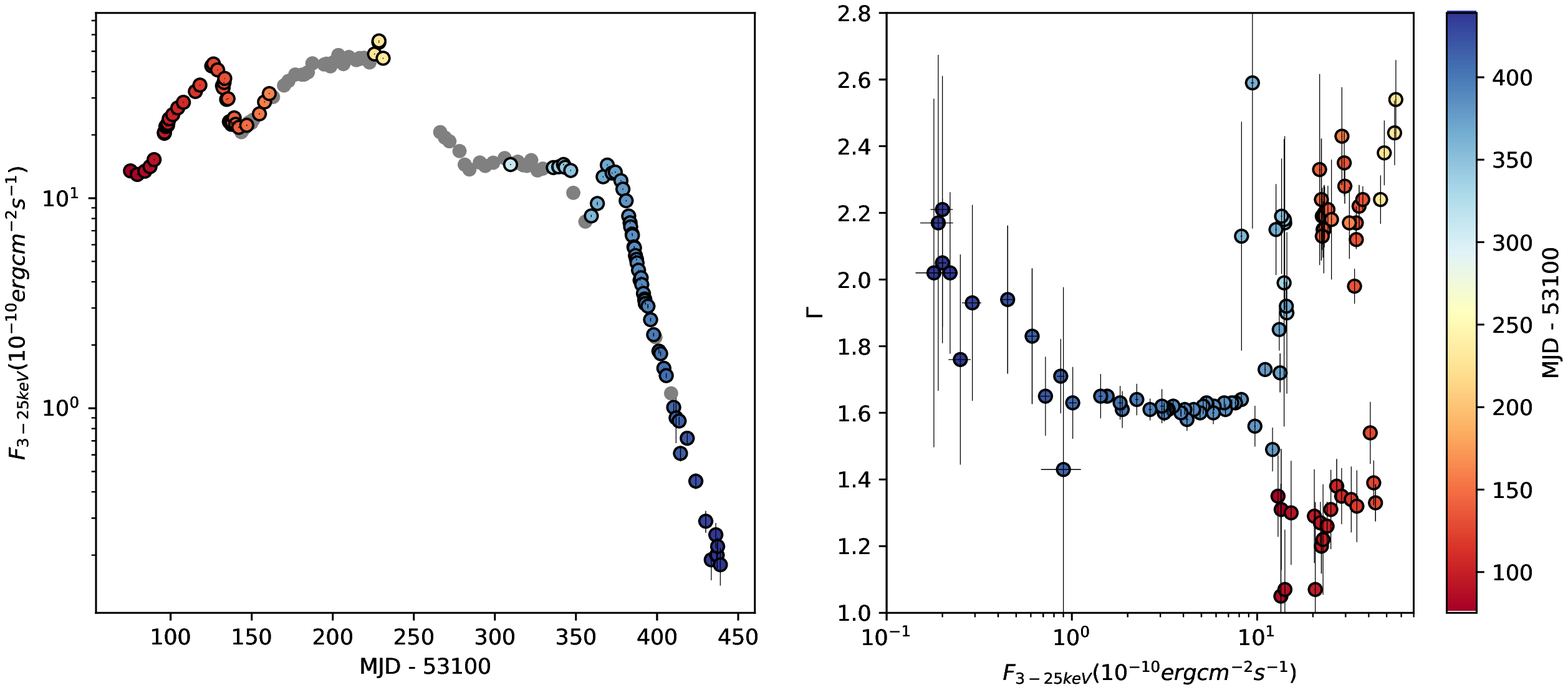}
\caption{Same as Figure 1, but for outburst 2004-2005 of GX 339-4.
}
\label{fig:2009p2013}
\end{figure*}

\begin{figure*}
\centering
\includegraphics[width=17cm,height=6.2cm]{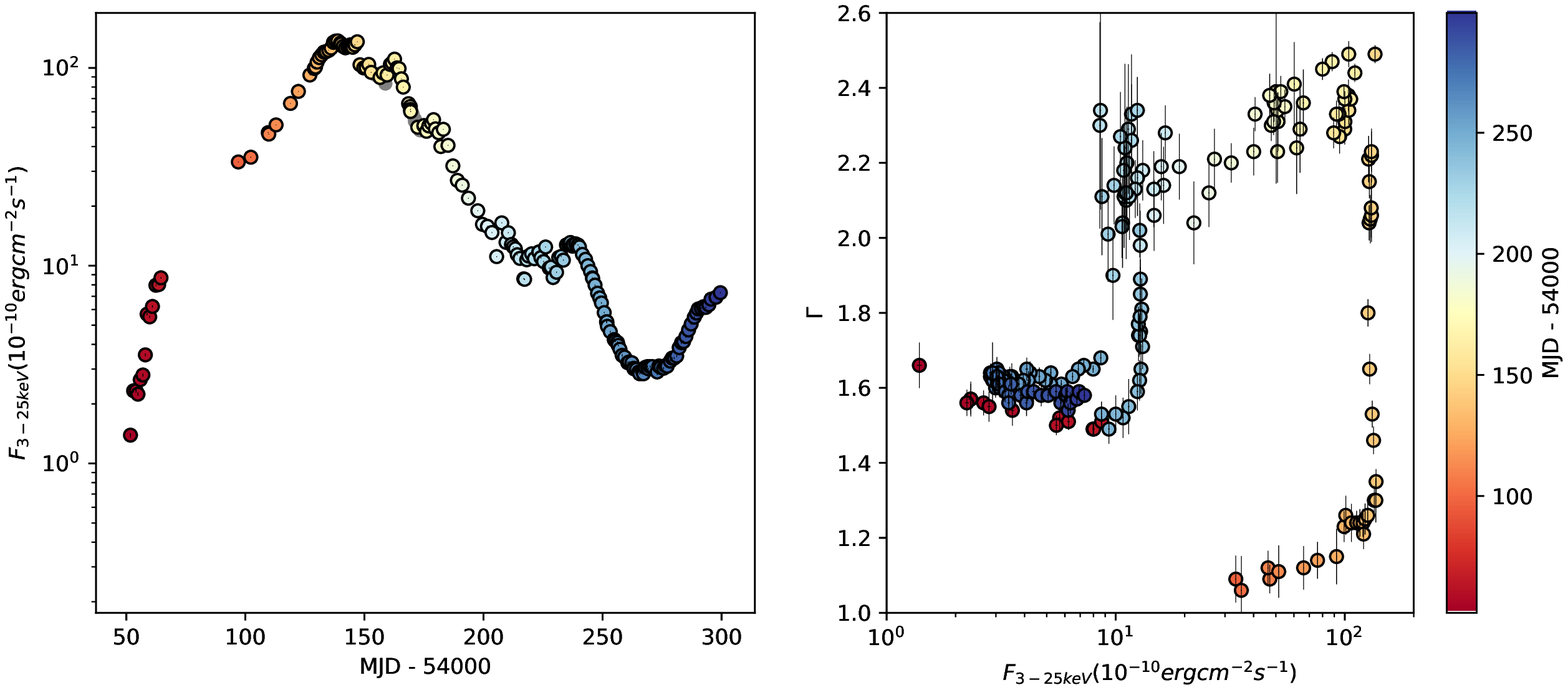}
\caption{Same as Figure 1, but for outburst 2006-2007 of GX 339-4.
}
\label{fig:2009p2013}
\end{figure*}

\begin{figure*}
\centering
\includegraphics[width=17cm,height=6.2cm]{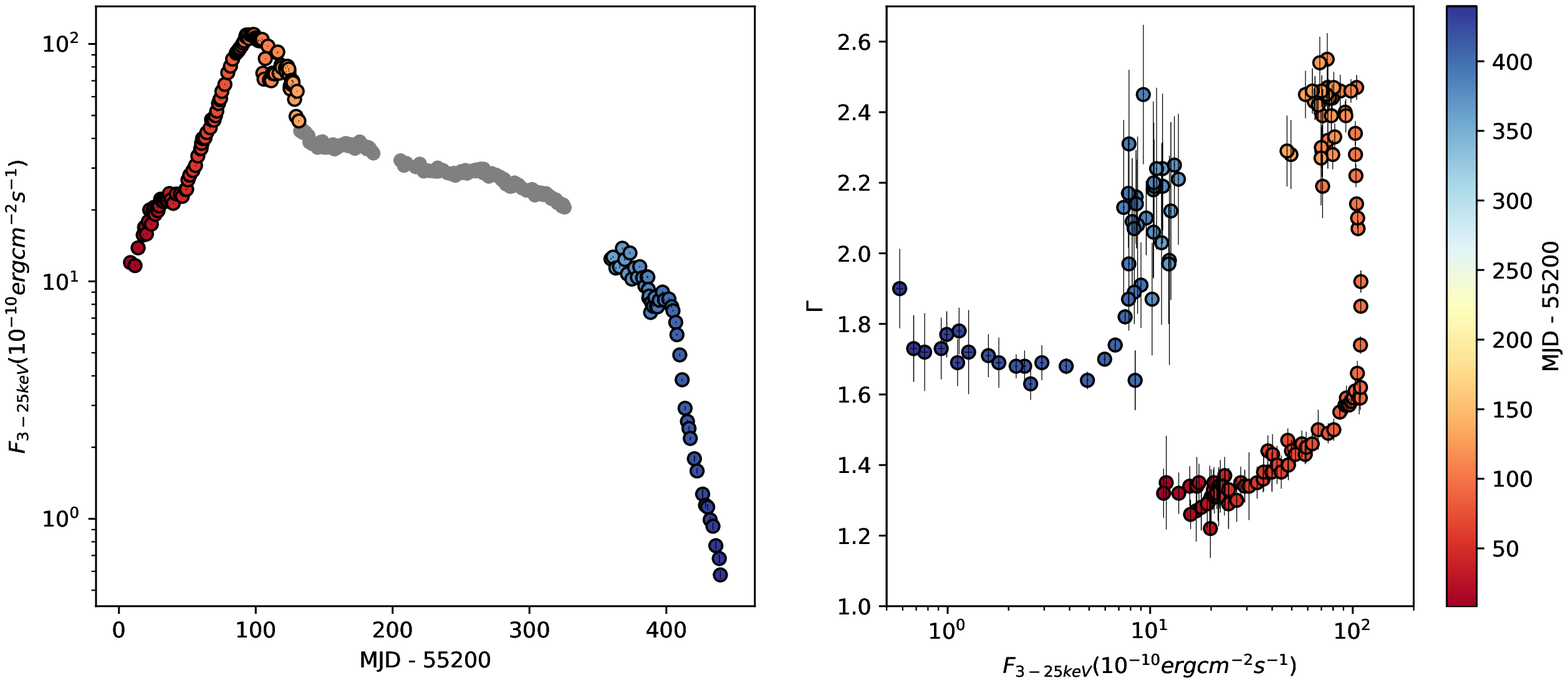}
\caption{Same as Figure 1, but for outburst 2010-2011 of GX 339-4. It should be cautious that only PCA data are adopted in this outburst due to HEXTE is ceased working.
}
\label{fig:2009p2013}
\end{figure*}

\begin{figure*}
\centering
\includegraphics[width=17cm,height=6.2cm]{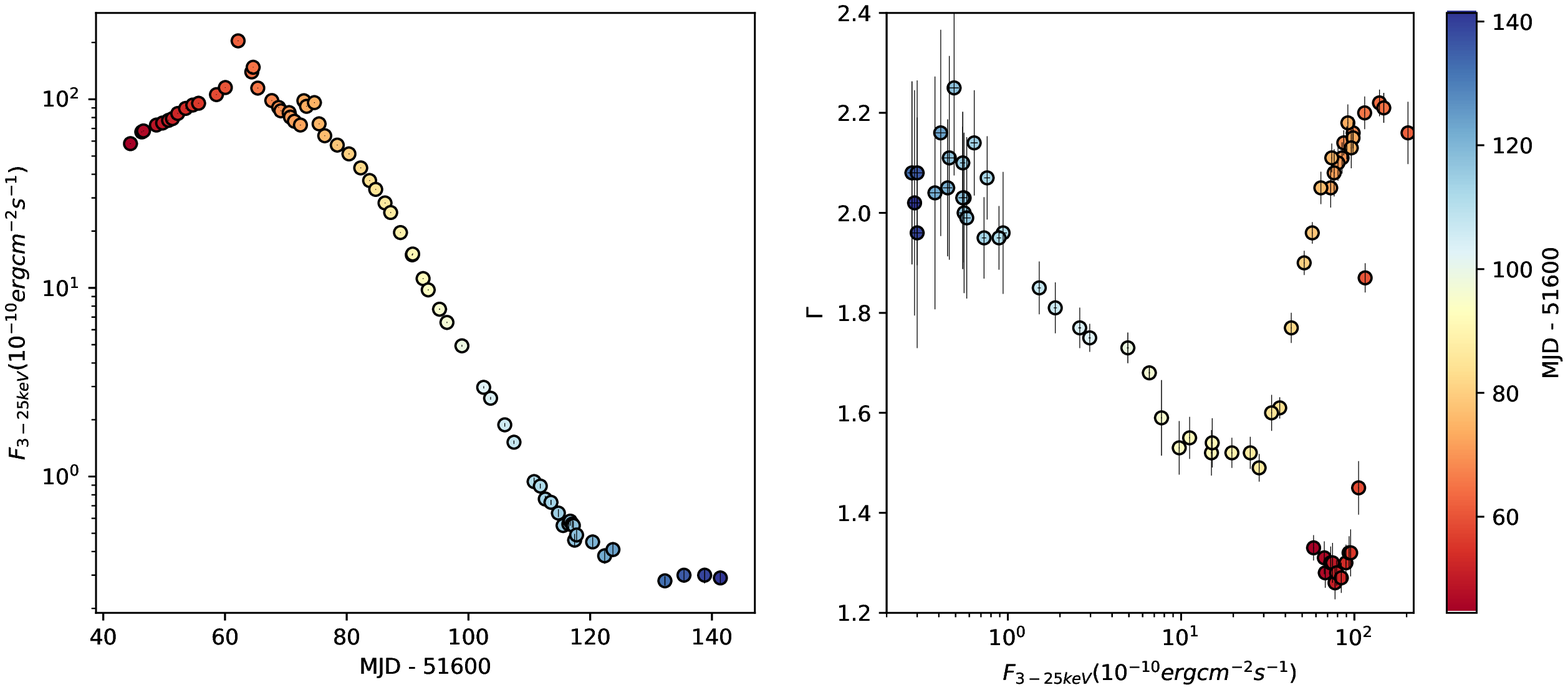}
\caption{Same as Figure 1, but for outburst 2000 of XTE J1550-564.
}
\label{fig:2009p2013}
\end{figure*}

\begin{figure*}
\centering
\includegraphics[width=17cm,height=6.2cm]{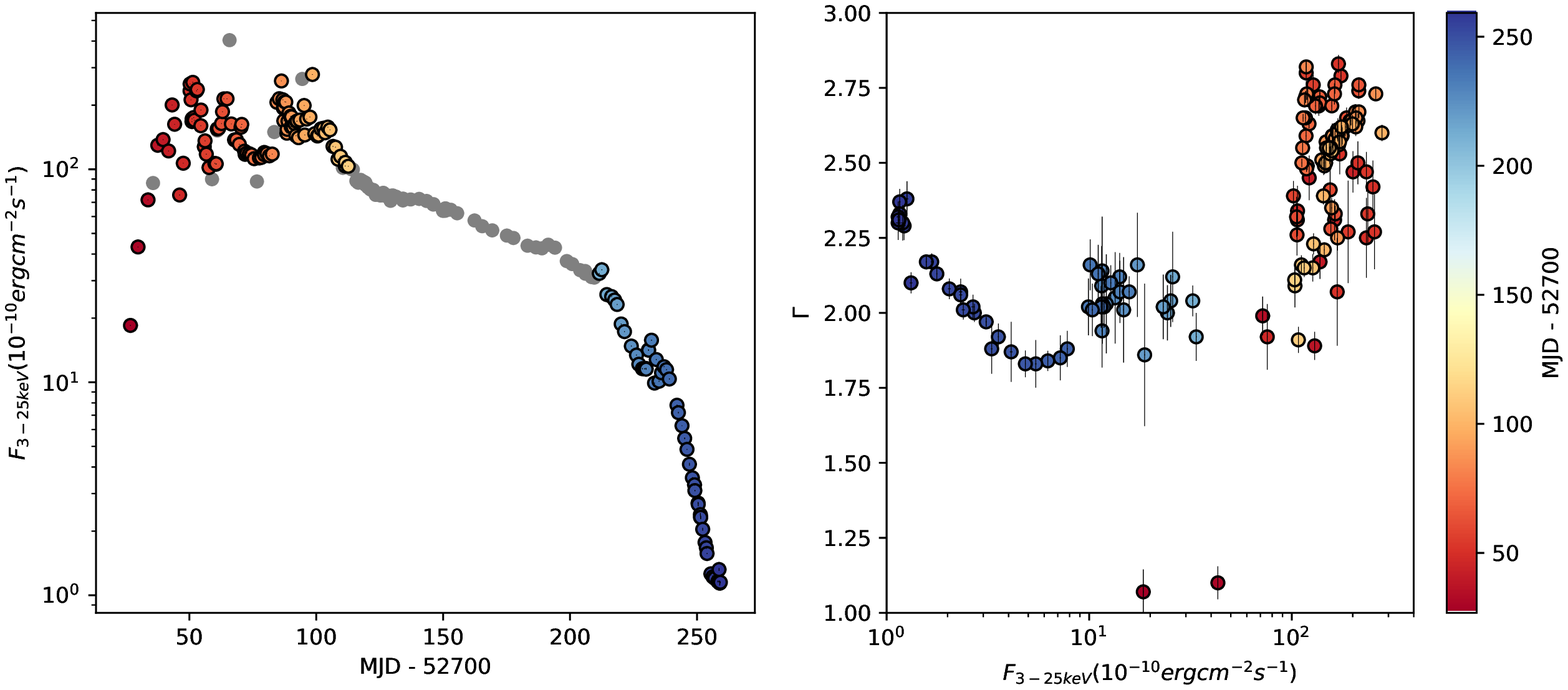}
\caption{Same as Figure 1, but for outburst 2003 of H1743-322.
}
\label{fig:2009p2013}
\end{figure*}

\begin{figure*}
\centering
\includegraphics[width=17cm,height=6.2cm]{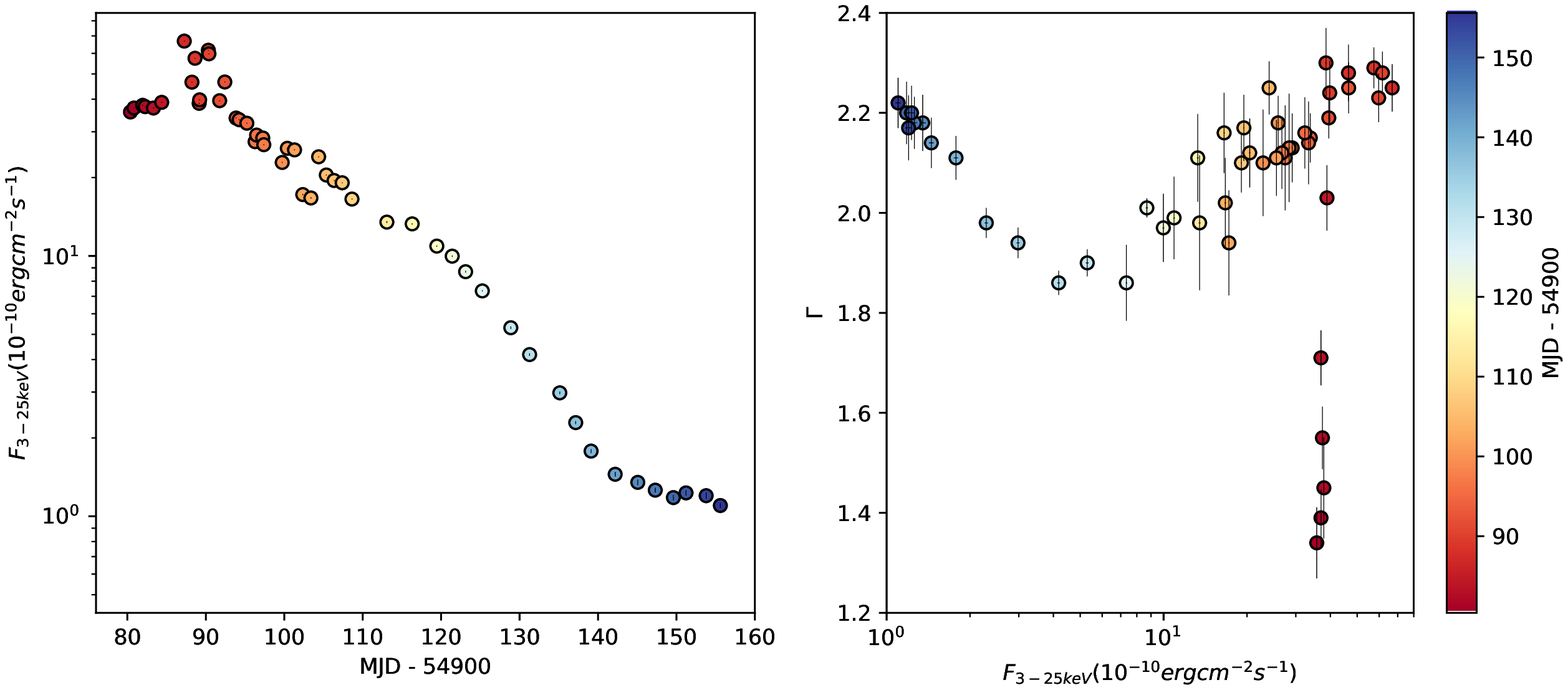}
\caption{Same as Figure 1, but for outburst 2009 of H1743-322.
}
\label{fig:2009p2013}
\end{figure*}

\begin{figure*}
\centering
\includegraphics[width=17cm,height=6.2cm]{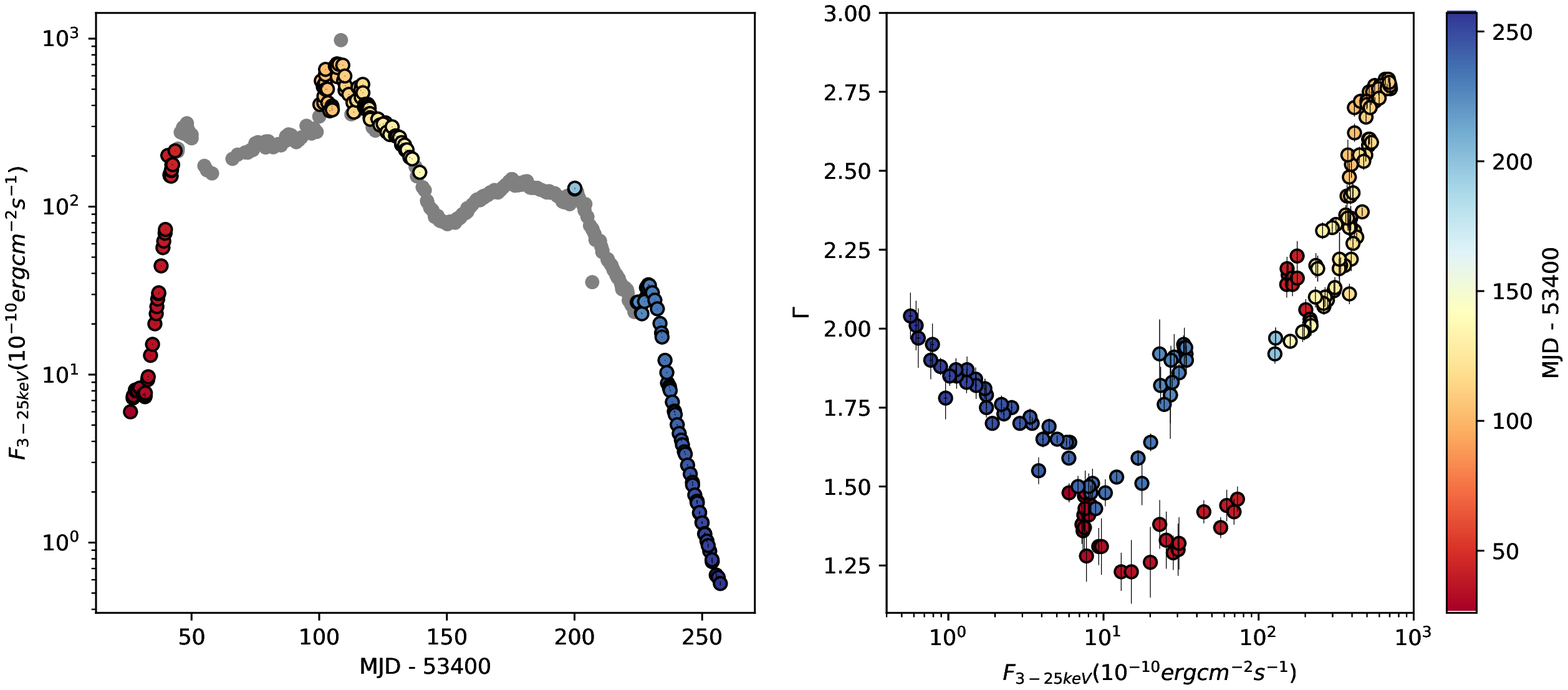}
\caption{Same as Figure 1, but for outburst 2005 of GRO J1655-40. 
}
\label{fig:2009p2013}
\end{figure*}

\begin{figure*}
\centering
\includegraphics[width=11cm,height=7.5cm]{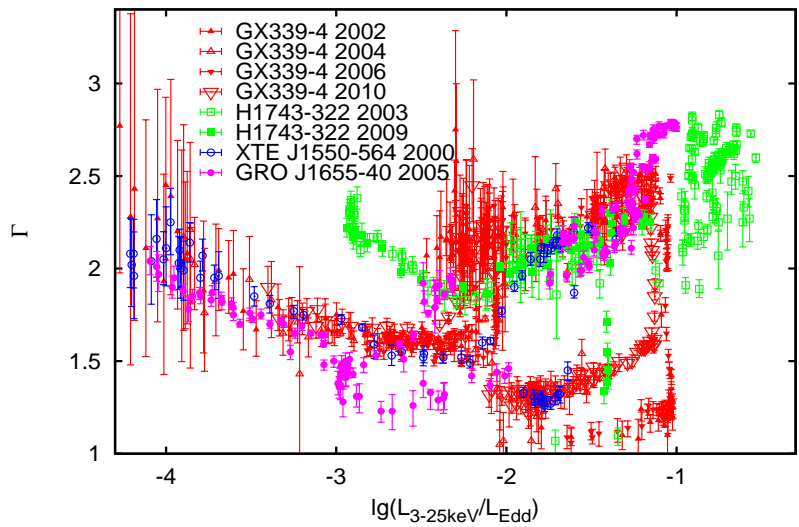}
\caption{The relation between $\Gamma$ and $L_{\rm 3-25keV}/L_{\rm Edd}$ for all eight outbursts. The inverted open triangles represent the outburst of GX 339-4 in 2010-2011, where only PCA data are considered due to the absence of HEXTE data.}
\label{fig:2009p2013}
\end{figure*}

\newpage

\begin{table*}
\centering
\caption{Data of outburst 2002-2003 for GX 339-4\label{tab:pks1424p240}.}
\begin{tabular}{lllllllll}
\hline \hline
  Obs. &Date & Diskbb & Diskbb & Power-Law & Power-Law & Flux & $\chi$$^{2}$ & Model \\
    & (MJD) & $T_{\rm in}$(keV) & Norm. & Index($\Gamma$) & Norm. & (3-25keV)  & /DOF &  \\
  ~[1]&[2] &[3] &[4] &[5] &[6] &[7] &[8] &[9] \\
  \hline
  60705-01-48-00 & 52324.39 & ... & ... & $2.11_{-0.59}^{+0.69}$ & $0.004_{-0.004}^{+0.01}$ & $0.11_{-0.03}^{+0.03}$ & 0.93 & 2 \\
  60705-01-49-00 & 52332.20 & ... & ... & $2.39_{-0.53}^{+0.63}$ & $0.01_{-0.01}^{+0.02}$ & $0.16_{-0.03}^{+0.03}$ & 0.84 & 2 \\
  60705-01-51-00 & 52339.11 & ... & ... & $2.45_{-0.41}^{+0.47}$ & $0.01_{-0.01}^{+0.01}$ & $0.15_{-0.02}^{+0.02}$ & 0.65 & 2 \\
  60705-01-52-00 & 52345.05 & ... & ... & $2.05_{-0.37}^{+0.42}$ & $0.01_{-0.003}^{+0.01}$ & $0.19_{-0.03}^{+0.03}$ & 0.85 & 2 \\
  60705-01-53-00 & 52350.07 & ... & ... & $1.99_{-0.21}^{+0.23}$ & $0.01_{-0.003}^{+0.004}$ & $0.30_{-0.02}^{+0.02}$ & 0.75 & 2 \\
  60705-01-54-00 & 52359.64 & ... & ... & $1.67_{-0.07}^{+0.07}$ & $0.02_{-0.002}^{+0.003}$ & $1.32_{-0.04}^{+0.04}$ & 0.70 & 2 \\
  60705-01-55-00 & 52366.60 & $1.74_{-0.16}^{+0.15}$ & $2.58_{-0.55}^{+0.78}$ & $1.14_{-0.12}^{+0.10}$ & $0.10_{-0.02}^{+0.03}$ & $23.61_{-0.13}^{+0.13}$ & 1.31 & 5 \\
  70109-01-02-00 & 52367.76 & $1.67_{-0.11}^{+0.09}$ & $5.08_{-0.65}^{+1.12}$ & $1.09_{-0.05}^{+0.05}$ & $0.13_{-0.01}^{+0.02}$ & $35.35_{-0.09}^{+0.09}$ & 0.84 & 5 \\
  70109-01-01-00 & 52371.54 & $1.65_{-0.20}^{+0.13}$ & $11.93_{-2.11}^{+5.82}$ & $1.10_{-0.08}^{+0.07}$ & $0.28_{-0.05}^{+0.06}$ & $70.56_{-0.21}^{+0.22}$ & 0.69 & 5 \\
  70110-01-01-00 & 52373.24 & $1.70_{-0.31}^{+0.13}$ & $14.22_{-2.19}^{+12.44}$ & $1.08_{-0.10}^{+0.11}$ & $0.33_{-0.08}^{+0.10}$ & $87.67_{-0.30}^{+0.30}$ & 0.98 & 5 \\
 ...  & ...  & ...  &... & ...  & ...  & ...  &...  \\ 
  \hline
 \end{tabular}
 \begin{minipage}{170mm}
Column [1]: Observation IDs; Column [2]: MJD Time; Columns [3] and [4]: The temperature and normalization value of disk blackbody; Columns [5].and [6]: The photon index and normalization value of power-law component; Column [7]: 3-25keV flux (with the unit of $10^{-10}\rm erg cm^{-2} s^{-1}$); Column [8]: Reduced $\chi^{2}$; Column [9]: the model in fitting ( 1. diskbb+gau+pow; 2. pow; 3. gau+pow; 4. diskbb+pow; 5. diskbb+gau+cutoff)
\end{minipage}

\end{table*}

\begin{table*}
\centering
\caption{Data of outburst 2004-2005 for GX 339-4\label{tab:pks1424p240}.}
\begin{tabular}{lllllllll}
\hline \hline
  Obs. &Date & Diskbb & Diskbb & Power-Law & Power-Law & Flux & $\chi$$^{2}$ & Model \\
    & (MJD) & $T_{\rm in}$(keV) & Norm. & Index($\Gamma$) & Norm. & (3-25keV)  & /DOF &  \\
  ~[1]&[2] &[3] &[4] &[5] &[6] &[7] &[8] &[9] \\
  \hline
  80102-04-92-00 & 53175.25 & $1.61_{-0.40}^{+0.28}$ & $1.60_{-0.66}^{+2.46}$ & $1.31_{-0.18}^{+0.13}$ & $0.08_{-0.03}^{+0.03}$ & $13.48_{-0.09}^{+0.09}$ & 1.04 & 5 \\
  80102-04-93-00 & 53179.51 & $1.06_{-0.20}^{+0.34}$ & $9.82_{-6.60}^{+16.54}$ & $1.35_{-0.12}^{+0.10}$ & $0.08_{-0.02}^{+0.02}$ & $12.96_{-0.09}^{+0.09}$ & 0.94 & 5 \\
  80102-04-94-00 & 53184.10 & $1.62_{-0.18}^{+0.26}$ & $2.28_{-0.61}^{+0.94}$ & $1.05_{-0.34}^{+0.13}$ & $0.05_{-0.02}^{+0.02}$ & $13.43_{-0.09}^{+0.09}$ & 1.10 & 5 \\
  80102-04-95-00 & 53187.31 & $1.69_{-0.24}^{+0.18}$ & $2.29_{-0.55}^{+1.21}$ & $1.07_{-0.18}^{+0.15}$ & $0.05_{-0.02}^{+0.01}$ & $14.12_{-0.09}^{+0.09}$ & 0.69 & 5 \\
  80102-04-96-00 & 53189.83 & $1.62_{-0.37}^{+0.22}$ & $1.94_{-0.70}^{+2.44}$ & $1.30_{-0.16}^{+0.13}$ & $0.09_{-0.03}^{+0.03}$ & $15.25_{-0.10}^{+0.10}$ & 0.85 & 5 \\
  90418-01-01-04 & 53196.14 & $1.67_{-0.31}^{+0.14}$ & $2.59_{-0.37}^{+1.75}$ & $1.29_{-0.14}^{+0.12}$ & $0.11_{-0.03}^{+0.05}$ & $20.36_{-0.17}^{+0.17}$ & 0.89 & 5 \\
  90418-01-01-00 & 53196.20 & $1.88_{-0.21}^{+0.17}$ & $2.57_{-0.58}^{+0.81}$ & $1.07_{-0.26}^{+0.21}$ & $0.07_{-0.03}^{+0.05}$ & $20.59_{-0.16}^{+0.16}$ & 0.74 & 5 \\
  90418-01-01-01 & 53197.09 & $1.31_{-0.15}^{+0.23}$ & $7.88_{-3.44}^{+5.32}$ & $1.27_{-0.06}^{+0.05}$ & $0.12_{-0.02}^{+0.02}$ & $21.93_{-0.07}^{+0.07}$ & 1.24 & 5 \\
  90418-01-01-02 & 53198.13 & $1.62_{-0.19}^{+0.12}$ & $3.94_{-0.64}^{+1.71}$ & $1.20_{-0.09}^{+0.08}$ & $0.10_{-0.02}^{+0.02}$ & $22.22_{-0.07}^{+0.07}$ & 0.81 & 5 \\
 ...  & ...  & ...  &... & ...  & ...  & ...  &...  \\ 
  \hline  
 \end{tabular}
\end{table*}

\begin{table*}
\centering
\caption{Data of outburst 2006-2007 for GX 339-4\label{tab:pks1424p240}.}
\begin{tabular}{lllllllll}
\hline \hline
  Obs. &Date & Diskbb & Diskbb & Power-Law & Power-Law & Flux & $\chi$$^{2}$ & Model \\
    & (MJD) & $T_{\rm in}$(keV) & Norm. & Index($\Gamma$) & Norm. & (3-25keV)  & /DOF &  \\
  ~[1]&[2] &[3] &[4] &[5] &[6] &[7] &[8] &[9] \\
  \hline
  91105-04-17-10 & 54051.74 & ... & ... & $1.66_{-0.06}^{+0.06}$ & $0.02_{-0.002}^{+0.002}$ & $1.39_{-0.04}^{+0.04}$ & 0.78 & 2 \\
  92052-07-01-00 & 54053.05 & ... & ... & $1.57_{-0.04}^{+0.04}$ & $0.03_{-0.002}^{+0.002}$ & $2.33_{-0.04}^{+0.04}$ & 0.85 & 2 \\
  92052-07-01-01 & 54054.10 & ... & ... & $1.57_{-0.04}^{+0.04}$ & $0.03_{-0.002}^{+0.002}$ & $2.33_{-0.04}^{+0.04}$ & 0.86 & 2 \\
  92052-07-01-02 & 54054.88 & ... & ... & $1.56_{-0.04}^{+0.04}$ & $0.02_{-0.002}^{+0.002}$ & $2.24_{-0.04}^{+0.04}$ & 1.14 & 2 \\
  92052-07-01-03 & 54055.86 & ... & ... & $1.56_{-0.03}^{+0.03}$ & $0.03_{-0.002}^{+0.002}$ & $2.65_{-0.04}^{+0.04}$ & 0.90 & 2 \\
  92052-07-02-00 & 54056.97 & ... & ... & $1.55_{-0.04}^{+0.04}$ & $0.03_{-0.002}^{+0.003}$ & $2.80_{-0.05}^{+0.05}$ & 1.19 & 2 \\
  92052-07-02-01 & 54057.99 & ... & ... & $1.54_{-0.04}^{+0.04}$ & $0.04_{-0.003}^{+0.003}$ & $3.54_{-0.06}^{+0.06}$ & 1.14 & 2 \\
  92052-07-02-02 & 54058.74 & ... & ... & $1.52_{-0.03}^{+0.03}$ & $0.06_{-0.003}^{+0.003}$ & $5.69_{-0.07}^{+0.07}$ & 1.30 & 2 \\
  92052-07-02-03 & 54059.82 & ... & ... & $1.50_{-0.03}^{+0.03}$ & $0.05_{-0.003}^{+0.003}$ & $5.51_{-0.07}^{+0.07}$ & 1.02 & 2 \\
  92052-07-02-04 & 54060.97 & ... & ... & $1.51_{-0.03}^{+0.03}$ & $0.06_{-0.003}^{+0.003}$ & $6.22_{-0.07}^{+0.07}$ & 0.91 & 2 \\
 ...  & ...  & ...  &... & ...  & ...  & ...  &...  \\ 
  \hline
\end{tabular}
\end{table*}

\begin{table*}
\centering
\caption{Data of outburst 2010-2011 for GX 339-4\label{tab:pks1424p240}.}
\begin{tabular}{lllllllll}
\hline \hline
  Obs. &Date & Diskbb & Diskbb & Power-Law & Power-Law & Flux & $\chi$$^{2}$ & Model \\
    & (MJD) & $T_{\rm in}$(keV) & Norm. & Index($\Gamma$) & Norm. & (3-25keV)  & /DOF &  \\
  ~[1]&[2] &[3] &[4] &[5] &[6] &[7] &[8] &[9] \\
  \hline
  95409-01-01-00 & 55208.48 & $1.51_{-0.50}^{+0.51}$ & $1.17_{-0.72}^{+4.38}$ & $1.35_{-0.13}^{+0.09}$ & $0.07_{-0.02}^{+0.02}$ & $12.00_{-0.11}^{+0.11}$ & 0.65 & 1 \\
  95409-01-02-00 & 55211.81 & $1.68_{-0.30}^{+0.23}$ & $0.92_{-0.30}^{+0.63}$ & $1.32_{-0.07}^{+0.07}$ & $0.07_{-0.01}^{+0.01}$ & $11.65_{-0.07}^{+0.07}$ & 0.97 & 1 \\
  95409-01-02-01 & 55214.08 & $1.48_{-0.25}^{+0.23}$ & $1.80_{-0.64}^{+1.71}$ & $1.32_{-0.06}^{+0.05}$ & $0.08_{-0.01}^{+0.01}$ & $13.84_{-0.06}^{+0.06}$ & 0.81 & 1 \\
  95409-01-02-02 & 55217.81 & $1.62_{-0.38}^{+0.19}$ & $1.36_{-0.37}^{+2.07}$ & $1.34_{-0.05}^{+0.05}$ & $0.09_{-0.01}^{+0.01}$ & $15.68_{-0.06}^{+0.06}$ & 0.59 & 1 \\
  95409-01-03-02 & 55218.73 & $1.57_{-0.31}^{+0.23}$ & $2.25_{-0.72}^{+2.48}$ & $1.27_{-0.08}^{+0.07}$ & $0.08_{-0.02}^{+0.01}$ & $16.86_{-0.10}^{+0.10}$ & 0.52 & 1 \\
  95409-01-03-03 & 55219.46 & $1.67_{-0.38}^{+0.25}$ & $1.29_{-0.48}^{+1.26}$ & $1.34_{-0.09}^{+0.07}$ & $0.10_{-0.02}^{+0.02}$ & $16.99_{-0.11}^{+0.11}$ & 0.69 & 1 \\
  95409-01-03-00 & 55220.17 & $1.84_{-0.39}^{+0.15}$ & $1.20_{-0.27}^{+1.05}$ & $1.26_{-0.04}^{+0.09}$ & $0.08_{-0.02}^{+0.02}$ & $15.83_{-0.09}^{+0.09}$ & 0.60 & 1 \\
  95409-01-03-04 & 55221.55 & $1.59_{-0.27}^{+0.20}$ & $2.25_{-0.64}^{+1.89}$ & $1.28_{-0.07}^{+0.06}$ & $0.09_{-0.02}^{+0.02}$ & $17.87_{-0.09}^{+0.09}$ & 0.77 & 1 \\
  95409-01-03-05 & 55222.57 & $1.72_{-0.25}^{+0.20}$ & $1.67_{-0.46}^{+1.56}$ & $1.31_{-0.08}^{+0.07}$ & $0.11_{-0.02}^{+0.02}$ & $20.00_{-0.11}^{+0.11}$ & 0.60 & 1 \\
  95409-01-03-01 & 55223.76 & $1.41_{-0.29}^{+0.26}$ & $2.89_{-1.18}^{+4.39}$ & $1.35_{-0.06}^{+0.04}$ & $0.11_{-0.02}^{+0.01}$ & $17.42_{-0.08}^{+0.08}$ & 0.83 & 1 \\
 ...  & ...  & ...  &... & ...  & ...  & ...  &...  \\ 
  \hline
 \end{tabular}
\end{table*}
  
\begin{table*}
\centering
\caption{Data of outburst 2000 for XTE J1550-564\label{tab:pks1424p240}.}
\begin{tabular}{lllllllll}
\hline \hline
  Obs. &Date & Diskbb & Diskbb & Power-Law & Power-Law & Flux & $\chi$$^{2}$ & Model \\
    & (MJD) & $T_{\rm in}$(keV) & Norm. & Index($\Gamma$) & Norm. & (3-25keV)  & /DOF &  \\
  ~[1]&[2] &[3] &[4] &[5] &[6] &[7] &[8] &[9] \\
  \hline
  50137-02-01-00 & 51644.47 & $0.98_{-0.07}^{+0.09}$ & $80.36_{-27.76}^{+40.07}$ & $1.33_{-0.03}^{+0.03}$ & $0.38_{-0.02}^{+0.02}$ & $58.16_{-0.14}^{+0.15}$ & 1.09 & 5 \\
  50137-02-02-00 & 51646.33 & $0.99_{-0.07}^{+0.08}$ & $95.57_{-32.21}^{+46.30}$ & $1.31_{-0.03}^{+0.03}$ & $0.41_{-0.03}^{+0.03}$ & $67.07_{-0.19}^{+0.19}$ & 0.72 & 5 \\
  50137-02-02-01 & 51646.61 & $1.02_{-0.07}^{+0.08}$ & $86.19_{-27.52}^{+39.03}$ & $1.28_{-0.03}^{+0.03}$ & $0.39_{-0.03}^{+0.03}$ & $67.93_{-0.18}^{+0.18}$ & 1.09 & 5 \\
  50137-02-03-00 & 51648.73 & $1.02_{-0.07}^{+0.08}$ & $91.13_{-29.14}^{+41.02}$ & $1.30_{-0.03}^{+0.03}$ & $0.44_{-0.03}^{+0.03}$ & $72.96_{-0.19}^{+0.19}$ & 0.88 & 5 \\
  50137-02-03-01G & 51649.75 & $1.00_{-0.08}^{+0.09}$ & $99.27_{-35.37}^{+52.53}$ & $1.30_{-0.04}^{+0.04}$ & $0.45_{-0.04}^{+0.04}$ & $74.84_{-0.23}^{+0.23}$ & 0.96 & 5 \\
  50137-02-04-00 & 51650.73 & $1.03_{-0.07}^{+0.08}$ & $97.44_{-29.14}^{+40.37}$ & $1.26_{-0.03}^{+0.03}$ & $0.42_{-0.03}^{+0.03}$ & $77.30_{-0.20}^{+0.20}$ & 1.14 & 5 \\
  50137-02-04-01 & 51651.37 & $1.03_{-0.07}^{+0.08}$ & $95.59_{-28.65}^{+39.35}$ & $1.28_{-0.03}^{+0.03}$ & $0.46_{-0.03}^{+0.03}$ & $78.93_{-0.18}^{+0.19}$ & 0.89 & 5 \\
  50137-02-05-00 & 51652.24 & $0.99_{-0.07}^{+0.08}$ & $126.93_{-38.95}^{+54.47}$ & $1.27_{-0.03}^{+0.03}$ & $0.48_{-0.04}^{+0.04}$ & $84.04_{-0.22}^{+0.22}$ & 0.76 & 5 \\
  50137-02-05-01 & 51653.53 & $0.98_{-0.07}^{+0.07}$ & $142.28_{-43.84}^{+61.34}$ & $1.30_{-0.03}^{+0.03}$ & $0.54_{-0.04}^{+0.04}$ & $89.38_{-0.23}^{+0.23}$ & 0.82 & 5 \\
  50137-02-06-00 & 51654.71 & $1.01_{-0.07}^{+0.08}$ & $124.77_{-39.44}^{+55.06}$ & $1.32_{-0.03}^{+0.03}$ & $0.59_{-0.04}^{+0.04}$ & $93.17_{-0.22}^{+0.22}$ & 0.91 & 5 \\
 ...  & ...  & ...  &... & ...  & ...  & ...  &...  \\ 
  \hline
  \end{tabular}
\end{table*}

\begin{table*}
\centering
\caption{Data of outburst 2003 for H1743-322\label{tab:pks1424p240}.}
\begin{tabular}{lllllllll}
\hline \hline
  Obs. &Date & Diskbb & Diskbb & Power-Law & Power-Law & Flux & $\chi$$^{2}$ & Model \\
    & (MJD) & $T_{\rm in}$(keV) & Norm. & Index($\Gamma$) & Norm. & (3-25keV)  & /DOF &  \\
  ~[1]&[2] &[3] &[4] &[5] &[6] &[7] &[8] &[9] \\
  \hline
  80138-01-01-00 & 52726.80 & $1.51_{-0.20}^{+0.19}$ & $2.69_{-0.82}^{+1.81}$ & $1.07_{-0.07}^{+0.06}$ & $0.07_{-0.01}^{+0.01}$ & $18.49_{-0.07}^{+0.07}$ & 0.87 & 5 \\
  80138-01-02-00G & 52729.79 & $1.30_{-0.16}^{+0.18}$ & $9.24_{-3.41}^{+6.79}$ & $1.10_{-0.05}^{+0.04}$ & $0.22_{-0.02}^{+0.02}$ & $43.18_{-0.12}^{+0.12}$ & 1.17 & 5 \\
  80138-01-03-00G & 52733.73 & $0.94_{-0.24}^{+0.24}$ & $39.81_{-31.53}^{+116.90}$ & $1.99_{-0.06}^{+0.06}$ & $2.54_{-0.26}^{+0.27}$ & $71.83_{-0.18}^{+0.18}$ & 1.83 & 5 \\
  80138-01-05-00G & 52737.55 & $0.87_{-0.19}^{+0.19}$ & $123.43_{-84.71}^{+320.24}$ & $1.89_{-0.05}^{+0.05}$ & $4.25_{-0.36}^{+0.36}$ & $129.66_{-0.30}^{+0.29}$ & 1.47 & 5 \\
  80138-01-06-00 & 52739.65 & $0.92_{-0.11}^{+0.12}$ & $188.02_{-92.21}^{+181.09}$ & $2.17_{-0.05}^{+0.05}$ & $7.14_{-0.66}^{+0.62}$ & $137.86_{-0.32}^{+0.31}$ & 1.36 & 5 \\
  80138-01-07-00 & 52741.83 & $1.16_{-0.04}^{+0.04}$ & $164.75_{-28.14}^{+32.66}$ & $2.45_{-0.08}^{+0.07}$ & $8.89_{-1.17}^{+1.27}$ & $121.94_{-0.30}^{+0.30}$ & 1.28 & 5 \\
  80146-01-01-00 & 52743.22 & $1.23_{-0.03}^{+0.03}$ & $239.97_{-31.74}^{+34.79}$ & $2.47_{-0.07}^{+0.06}$ & $11.15_{-2.50}^{+2.58}$ & $200.57_{-0.47}^{+0.47}$ & 1.44 & 5 \\
  80146-01-02-00 & 52744.20 & $1.12_{-0.05}^{+0.04}$ & $245.47_{-42.87}^{+61.22}$ & $2.31_{-0.07}^{+0.07}$ & $9.92_{-1.23}^{+1.31}$ & $162.76_{-0.38}^{+0.38}$ & 1.00 & 5 \\
  80146-01-03-00 & 52746.17 & $1.12_{-0.04}^{+0.04}$ & $156.66_{-22.86}^{+28.06}$ & $1.92_{-0.11}^{+0.10}$ & $2.25_{-0.41}^{+0.46}$ & $75.72_{-0.20}^{+0.21}$ & 1.54 & 5 \\
  80146-01-03-01 & 52747.61 & $1.14_{-0.03}^{+0.03}$ & $193.40_{-26.32}^{+26.90}$ & $2.34_{-0.08}^{+0.08}$ & $6.34_{-0.88}^{+1.01}$ & $106.79_{-0.27}^{+0.27}$ & 1.25 & 5 \\
 ...  & ...  & ...  &... & ...  & ...  & ...  &...  \\ 
  \hline
  \end{tabular}
\end{table*}

\begin{table*}
\centering
\caption{Data of outburst 2009 for H1743-322\label{tab:pks1424p240}.}
\begin{tabular}{lllllllll}
\hline \hline
  Obs. &Date & Diskbb & Diskbb & Power-Law & Power-Law & Flux & $\chi$$^{2}$ & Model \\
    & (MJD) & $T_{\rm in}$(keV) & Norm. & Index($\Gamma$) & Norm. & (3-25keV)  & /DOF &  \\
  ~[1]&[2] &[3] &[4] &[5] &[6] &[7] &[8] &[9] \\
  \hline
  94413-01-02-00 & 54980.39 & $1.32_{-0.18}^{+0.29}$ & $9.74_{-4.83}^{+8.11}$ & $1.34_{-0.07}^{+0.07}$ & $0.26_{-0.04}^{+0.04}$ & $35.72_{-0.13}^{+0.13}$ & 0.94 & 5 \\
  94413-01-02-02 & 54980.84 & $1.33_{-0.17}^{+0.21}$ & $8.97_{-3.51}^{+7.02}$ & $1.39_{-0.06}^{+0.05}$ & $0.30_{-0.04}^{+0.03}$ & $37.00_{-0.11}^{+0.11}$ & 0.88 & 5 \\
  94413-01-02-01 & 54981.95 & $1.40_{-0.25}^{+0.24}$ & $7.61_{-2.96}^{+8.58}$ & $1.45_{-0.10}^{+0.08}$ & $0.35_{-0.07}^{+0.03}$ & $37.90_{-0.15}^{+0.15}$ & 0.88 & 5 \\
  94413-01-02-05 & 54982.27 & $1.10_{-0.16}^{+0.22}$ & $17.30_{-10.09}^{+19.23}$ & $1.55_{-0.06}^{+0.05}$ & $0.43_{-0.05}^{+0.05}$ & $37.44_{-0.12}^{+0.12}$ & 0.89 & 5 \\
  94413-01-02-04 & 54983.32 & $0.99_{-0.11}^{+0.14}$ & $39.56_{-19.50}^{+33.63}$ & $1.71_{-0.06}^{+0.05}$ & $0.61_{-0.07}^{+0.07}$ & $36.99_{-0.12}^{+0.12}$ & 1.13 & 5 \\
  94413-01-02-03 & 54984.37 & $0.88_{-0.06}^{+0.06}$ & $144.33_{-45.34}^{+64.27}$ & $2.03_{-0.07}^{+0.06}$ & $1.15_{-0.15}^{+0.15}$ & $38.90_{-0.13}^{+0.13}$ & 1.10 & 5 \\
  94413-01-03-00 & 54987.25 & $0.94_{-0.03}^{+0.03}$ & $534.53_{-72.11}^{+83.46}$ & $2.25_{-0.05}^{+0.04}$ & $2.23_{-0.27}^{+0.28}$ & $66.87_{-0.23}^{+0.23}$ & 1.33 & 1 \\
  94413-01-03-01 & 54988.23 & $0.89_{-0.02}^{+0.02}$ & $734.47_{-77.05}^{+86.56}$ & $2.28_{-0.05}^{+0.05}$ & $1.35_{-0.18}^{+0.20}$ & $46.51_{-0.17}^{+0.17}$ & 0.78 & 1 \\
  94413-01-03-07 & 54988.62 & $0.95_{-0.02}^{+0.02}$ & $514.05_{-57.81}^{+65.65}$ & $2.29_{-0.04}^{+0.03}$ & $1.94_{-0.19}^{+0.19}$ & $57.45_{-0.17}^{+0.16}$ & 0.82 & 1 \\
  ...  & ...  & ...  &... & ...  & ...  & ...  &...  \\ 
  \hline
\end{tabular}

\end{table*}

\begin{table*}
\centering
\caption{Data of outburst 2005 for GRO J1655-40\label{tab:pks1424p240}.}
\begin{tabular}{lllllllll}
\hline \hline
  Obs. &Date & Diskbb & Diskbb & Power-Law & Power-Law & Flux & $\chi$$^{2}$ & Model \\
    & (MJD) & $T_{\rm in}$(keV) & Norm. & Index($\Gamma$) & Norm. & (3-25keV)  & /DOF &  \\
  ~[1]&[2] &[3] &[4] &[5] &[6] &[7] &[8] &[9] \\
  \hline
  90428-01-01-00 & 53426.04 & $1.37_{-0.24}^{+0.22}$ & $0.58_{-0.25}^{+0.66}$ & $1.48_{-0.03}^{+0.03}$ & $0.05_{-0.004}^{+0.002}$ & $6.00_{-0.02}^{+0.02}$ & 1.02 & 1 \\
  90058-16-05-00 & 53427.02 & $1.78_{-0.34}^{+0.25}$ & $0.42_{-0.16}^{+0.37}$ & $1.38_{-0.07}^{+0.06}$ & $0.05_{-0.01}^{+0.01}$ & $7.26_{-0.04}^{+0.04}$ & 0.98 & 1 \\
  90428-01-01-01 & 53427.15 & $1.58_{-0.22}^{+0.20}$ & $0.62_{-0.23}^{+0.39}$ & $1.41_{-0.05}^{+0.05}$ & $0.05_{-0.01}^{+0.01}$ & $7.47_{-0.04}^{+0.04}$ & 1.30 & 1 \\
  90058-16-07-00 & 53427.94 & $1.61_{-0.23}^{+0.20}$ & $0.63_{-0.23}^{+0.39}$ & $1.42_{-0.05}^{+0.05}$ & $0.06_{-0.01}^{+0.01}$ & $8.09_{-0.04}^{+0.04}$ & 0.85 & 1 \\
  90428-01-01-03 & 53428.13 & $1.43_{-0.26}^{+0.25}$ & $1.04_{-0.46}^{+1.27}$ & $1.41_{-0.05}^{+0.04}$ & $0.06_{-0.01}^{+0.01}$ & $8.04_{-0.04}^{+0.04}$ & 0.82 & 1 \\
  90428-01-01-04 & 53428.85 & $1.52_{-0.24}^{+0.20}$ & $0.40_{-0.16}^{+0.32}$ & $1.44_{-0.02}^{+0.02}$ & $0.06_{-0.004}^{+0.004}$ & $8.01_{-0.03}^{+0.03}$ & 1.29 & 1 \\
  90428-01-01-02 & 53429.71 & $1.52_{-0.21}^{+0.19}$ & $0.69_{-0.23}^{+0.49}$ & $1.44_{-0.03}^{+0.03}$ & $0.06_{-0.005}^{+0.005}$ & $8.31_{-0.03}^{+0.03}$ & 0.97 & 1 \\
  90428-01-01-05 & 53430.96 & $1.39_{-0.24}^{+0.21}$ & $0.70_{-0.29}^{+0.68}$ & $1.47_{-0.03}^{+0.03}$ & $0.06_{-0.01}^{+0.01}$ & $7.54_{-0.03}^{+0.03}$ & 1.05 & 1 \\
  90058-16-06-00 & 53431.17 & $1.43_{-0.38}^{+0.31}$ & $0.85_{-0.45}^{+1.64}$ & $1.43_{-0.07}^{+0.06}$ & $0.06_{-0.01}^{+0.01}$ & $7.60_{-0.05}^{+0.06}$ & 0.74 & 1 \\
  90428-01-01-06 & 53431.61 & $1.36_{-0.47}^{+0.22}$ & $0.82_{-0.55}^{+4.16}$ & $1.48_{-0.07}^{+0.05}$ & $0.07_{-0.01}^{+0.01}$ & $7.63_{-0.05}^{+0.05}$ & 1.00 & 1 \\
 ...  & ...  & ...  &... & ...  & ...  & ...  &...  \\ 
  \hline
 \end{tabular}
\end{table*}

\bsp	
\label{lastpage}

\begin{thebibliography}{}
\bibitem[Abramowicz et al.(1995)]{ab95} Abramowicz, M. A., Chen, X., Kato, S., Lasota, J.-P., \& Regev, O. 1995, \apjl, 438, L37

\bibitem[Allen et al.(2015)]{al15} Allen, J. L., Linares, M., Homan, J., Chakrabarty, D.  2015, \apj, 801, 10

\bibitem[Arnaud et al.(2017)]{arna17} Arnaud, K., Gordon, G., Dorman, B. https://heasarc.gsfc.nasa.gov/docs/software/lheasoft\\/xanadu/xspec/XspecManual.pdf

\bibitem[Balbus \& Henri(2008)]{bh08} Balbus S. A., Henri P., 2008, \apj, 674, 408

\bibitem[Beer  \& Podsiadlowski(2002)]{beer02} Beer, M. E., \& Podsiadlowski, P. 2002, MNRAS, 331, 351

\bibitem[Belloni et al.(2006)]{be06} Belloni T., et al., 2006, \mnras, 367, 1113

\bibitem[Begelman \& Armitage(2014)]{ba14} Begelman M. C., Armitage P. J., 2014, \apj, 782, L18

\bibitem[Cao(2009)]{cao09} Cao, X. 2009, \mnras, 394, 207

\bibitem[Cao et al.(2014)]{cao14} Cao, X.-F., Wu, Q., Dong, A.-J. 2014, \apj, 788, 52

\bibitem[Cao(2016)]{cao16} Cao, X., 2016, \apj, 817, 71

\bibitem[Coriat et al.(2011)]{cori11} Coriat, M., Corbel, S., Prat, L. et al. 2011, \mnras, 414, 677

\bibitem[Debnath et al.(2008)]{debn08} Debnath, D., Chakrabarti, Sandip K., Nandi, A. et al. 2008, BASI, 36, 151     

\bibitem[Dincer et al.(2012)]{dinc12} Dincer, T., Kalemeci, E., Buxton, M. M. et al. 2012, \apj, 753, 55

\bibitem[Done et al.(2010)]{done10} Done, C., \& Diaz Trigo, M. 2010, MNRAS, 407, 2287

\bibitem[Done et al.(2007)]{done07} Done, C., Gierlinski, M., Kubota, A. 2007, \aapr, 15, 1           

\bibitem[Dong \& Wu(2015)]{dong15} Dong, A.-J. \& Wu, Q. 2015, \mnras, 453, 3447

\bibitem[Dunn et al.(2010)]{dunn10} Dunn, R. J. H., Fender, R. P., K{\"o}rding, E. G. et al. 2010, \mnras, 403, 61

\bibitem[Eckersall et al.(2015)]{ecke15} Eckersall, A. J., Vaughan, S., Wynn, G. A. 2015, \mnras, 450, 3410

\bibitem[Emmanoulopoulos et al.(2012)]{em12} Emmanoulopoulos, D., Papadakis, I. E., McHardy, I. M., et al. 2012, \mnras, 424, 1327

\bibitem[Esin et al.(1997)]{esin97} Esin, A. A., McClintock, J. E., Narayan, R. 1997, \apj, 489, 865


\bibitem[Fender \& Belloni(2012)]{fend12} Fender, R. \& Belloni, T. 2012, Science, 337, 540                


\bibitem[Foellmi et al.(2006)]{foe06} Foellmi, C., Depagne, E., Dall, T. H., Mirabel, I. F. 2006, \aap, 457, 249

\bibitem[Gierlinski \& Netwon(2006)]{gier06} Gierlinski, M. \& Newton, J. 2006, \mnras, 370, 837

\bibitem[Gladstone et al.(2007)]{gl07} Gladstone, J., Done, C., Gierlinski, M. 2007, \mnras, 378, 13

\bibitem[Gliozzi et al.(2011)]{glio11} Gliozzi, M., Titarchuk, L., Satyapal, S. et al. 2011, \apj, 735, 16

\bibitem[Greene et al.(2001)]{gree01} Greene, J., Bailyn, C. D. \& Orosz, J. A. 2001, \apj, 554, 1290

\bibitem[Gu \& Cao(2009)]{gu09} Gu, M. \& Cao, X. 2009, \mnras, 399, 349

\bibitem[Heil et al.(2015)]{hei15} Heil, L. M., Uttley, P., Klein-Wolt, M. 2015, \mnras, 448, 3348

\bibitem[Hjalmarsdotter et al.(2008)]{hj08} Hjalmarsdotter, L., Zdziarski, A. A., Larsson, S., et al. 2008, \mnras, 384, 278

\bibitem[Hjellming \& Rupen(1995)]{hjel95} Hjellming, R. M. \& Rupen, M. P. 1995, Nature, 375, 464


\bibitem[Jang et al.(2014)]{jang14} Jang, I., Gliozzi, M., Hughes, C., Titarchuk, L. 2014, \mnras, 443, 72

\bibitem[Kalemci(2002)]{kale02} Kalemci, E.  2013, {\it Ph.D. Thesis}, University of California, San Diego

\bibitem[Kalemci et al.(2006)]{kale06} Kalemci, E., Tomsick, J. A., Rothschild, R. E. et al. 2006, \apj, 639, 340

\bibitem[Kalemci et al.(2013)]{kale13} Kalemci, E., Dincer, T., Tomsick, J. A. et al. 2013, \apj, 779, 95

\bibitem[Kalemci et al.(2016)]{kale16} Kalemci, E., Begelman, M. C., Maccarone, T. J. et al. 2016, \mnras, 463, 615

\bibitem[Kotov et al.(2001)]{koto01} Kotov, O., Churazov, E., Gilfanov, M. 2001, \mnras, 327, 799

\bibitem[Lasota(2001)]{laso01} Lasota, J.-P. 2001, \nar, 45, 449

\bibitem[Liu et al.(1999)]{liu99} Liu, B. F., Yuan, W., Meyer, F., Meyer-Hofmeister, E., \& Xie, G. Z. 1999, \apjl, 527, L17

\bibitem[Liu et al.(2005)]{liu05} Liu, B. F., Meyer, F., \& Meyer-Hofmeister, E. 2005, \aap, 442, 555

\bibitem[Liu et al.(2007)]{liu07} Liu, B.-F., Taam, R. E., Meyer-Hofmeister, E., \& Meyer, F. 2007, \apj, 671, 695

\bibitem[Liu et al.(2011)]{liu11} Liu, B. F., Done, C., \& Taam, R. E. 2011, \apj, 726, 10

\bibitem[Liu et al.(2016)]{liu16}  Liu, B.-F., Taam, R. E., Qiao, E., Yuan, W. 2016, IAUS, 312, 52

\bibitem[Maccarone \& Coppi(2003)]{ma03} Maccarone, T. J., \& Coppi, P. S. 2003, \mnras, 338, 189



\bibitem[Markoff et al.(2005)]{mar05} Markoff, S., Nowak, M. A., Wilms, J. 2005, ApJ, 635, 1203

\bibitem[Markoff et al.(2008)]{mar08} Markoff, S., et al. 2008, ApJ, 681, 905

\bibitem[Markowitzet al.(2009)]{mar09} Markowitz, A., Reeves, J. N., George, I. M., et al. 2009, \apj, 691, 922

\bibitem[Miller et al.(2006)]{mill06} Miller, J. M., Homan, J., \& Miniutti, G. 2006, \apj, 652, L113

\bibitem[Miller et al.(2012)]{mill12} Miller, J. M., Raymond, J., Fabian, A. C., et al. 2012, \apj, 759, 6M

\bibitem[Miyamoto et al.(1995)]{miy95} Miyamoto, S., Kitamoto, S., Hayashida, K., \& Egoshi, W. 1995, \apjl, 442, L13

\bibitem[Motta et al.(2009)]{mott09} Motta, S., Belloni, T., Homan, J. 2009, \mnras, 400, 1603

\bibitem[Motta et al.(2011)]{mott11} Motta, S., Mu$\tilde{n}$oz-Darias, T., Casella, P. et al. 2011, \mnras, 418, 2292

\bibitem[Motta et al.(2012)]{mott12} Motta, S., Homan, J., Mu$\tilde{n}$oz-Darias, T. et al. 2012, \mnras, 427, 595

\bibitem[Munoz-Darias et al.(2013)]{muno13} Mu$\tilde{n}$oz-Darias, T., Coriat, M., Plant, D. S. et al. 2013, \mnras, 432, 1330

\bibitem[Narayan \& McClintock(2008)]{nara08} Narayan, R. \& McClintock, J. E. 2008, \nar, 51, 733

\bibitem[Narayan(2005)]{nara05} Narayan, R. 2005, \apss, 300, 177

\bibitem[Narayan \& Yi(1995)]{ny95} Narayan, R., \& Yi, I. 1995, \apj, 452, 710

\bibitem[Nixon \& Salvesen(2014)]{ns14} Nixon C., Salvesen G., 2014, \mnras, 437, 3994

\bibitem[Orosz et al.(2002)]{oros02} Orosz, J. A., Groot, P. J., van der Klis, M. et al. 2002, \apj, 568, 845O

\bibitem[Orosz \& Bailyn(1997)]{oros97} Orosz, J. A., \& Bailyn, C. D. 1997, ApJ, 477, 876

\bibitem[Parker et al.(2016)]{park16} Parker, M. L., Tomsick, J. A., Kennea, J. A. 2016, ApJ, 821, 6

\bibitem[Plotkin et al.(2017)]{plot17} Plotkin, R. M., Miller-Jones, J. C. A., Gallo, E. et al. 2017, \apj, 834, 104

\bibitem[Plotkin et al.(2013)]{plot13} Plotkin, R. M.; Gallo, E.; Jonker, P. G. 2013, \apj, 773, 59

\bibitem[Qiao \& Liu(2010)]{qiao10} Qiao, E. \& Liu, B. F. 2010, \pasj, 62, 661                               

\bibitem[Qiao \& Liu(2013)]{qiao13} Qiao, E. \& Liu, B. F. 2013, \apj, 764, 2

\bibitem[Qiao \& Liu(2018)]{qiao18} Qiao, E. \& Liu, B. F. 2018, \mnras, 477, 210


\bibitem[Reig et al.(2013)]{reig13} Reig, P., Papadakis, I. E., Sobolewska, M. A. et al. 2013, \mnras, 435, 3395

\bibitem[Reis et al.(2008)]{reis08} Reis, R. C., Fabian, A. C., Ross, R. R., et al. 2008, MNRAS, 387, 1489


\bibitem[Remillard \& McClintock(2006)]{remi06} Remillard, R. A. \& McClintock, J. E. 2006, \araa, 44, 49

\bibitem[Shakura \& Sunyaev(1973)]{shak73} Shakura, N. I. \& Sunyaev, R. A.  1973, \aap, 24, 337

\bibitem[Shaposhnikov \& Titarchuk(2009)]{shap09} Shaposhnikov, N., Titarchuk, L., 2009, \apj, 699, 453

\bibitem[Shemmer et al.(2006)]{sh06} Shemmer, O., Brandt, W. N., Netzer, H., Maiolino, R., \& Kaspi, S. 2006, \apj, 646, 29

\bibitem[Sobolewska \& Papadakis(2009)]{sob09} Sobolewska, M. A., Papadakis, I. E. 2009, \mnras, 399, 1597

\bibitem[Steiner et al.(2016)]{stei16} Steiner, James F., Remillard, Ronald A., Garcia, Javier A.  2016, \apj, 829, 22

\bibitem[Stuchlik \& Kolos(2016)]{stuc16}  Stuchlik, Z., \& Kolos, M. 2016, \apj, 825, 13

\bibitem[Sunyaev \& Titarchuk(1980)]{suny80} Sunyaev, R. A. \& Titarchuk, L. G.  1980, \aap, 86, 121

\bibitem[Taam et al.(2018)]{tam18} Taam, R. E., Qiao, E., Liu, B. F., Meyer-Hofmeister, E. 2018, \apj, 860, 166

\bibitem[Titarchuk et al.(1997)]{tita97} Titarchuk, L., Mastichiadis, A., Kylafis, N. D.  1997, \apj, 487, 834

\bibitem[Titarchuk \& Shrader(2002)]{tita02} Titarchuk, L. \& Shrader, C. R.  2002, \apj, 567, 1057

\bibitem[Trichas et al.(2013)]{tr13} Trichas, M., Green, P. J., Constantin, A., et al. 2013, \apj, 778, 188


\bibitem[Wang et al.(2004)]{wang04} Wang, J.-M., Watarai, K.-Y., Mineshige, S., 2004, \apj, 607, 107

\bibitem[Weng et al.(2013)]{weng13}  Weng, S.-S., Zhang, S.-N., Ge, M.-Y., Li, J., Zhang, S. 2013, \apj, 763, 34

\bibitem[Weng \& Zhang(2011)]{weng11} Weng, S.-S., Zhang, S.-N., 2011, \apj, 739, 42

\bibitem[Wu et al.(2007)]{wu07} Wu Q., Yuan F., Cao X., 2007, \apj, 669, 96

\bibitem[Wu \& Gu(2008)]{wu08} Wu, Q. \& Gu, M.  2008, \apj, 682, 212

\bibitem[You et al.(2012)]{you12} You, B., Cao, X., \& Yuan, Y.-F. 2012, \apj, 761, 109

\bibitem[Yang et al.(2015)]{yang15} Yang, Q.-X., Xie, F.-G., Yuan, F. et al. 2015, \mnras, 447, 1692

\bibitem[Yu \& Yan(2009)]{yu09} Yu, W., \& Yan, Z. 2009, \apj, 701, 1940

\bibitem[Yan \& Yu(2015)]{yan15} Yan, Z. \& Yu, W. 2015, \apj, 805, 87

\bibitem[Yuan \& Cui(2005)]{yc05} Yuan F., Cui W., 2005, \apj, 629, 408

\bibitem[Yuan et al.(2007)]{yu07} Yuan F., Taam R. E., Misra R., Wu X.-B., Xue Y., 2007, \apj, 658, 282

\bibitem[Yuan et al.(2004)]{yu04} Yu, W., van der Klis, M., \& Fender, R. 2004, ApJL, 611, L121

\bibitem[Yuan et al.(2009)]{yuan09}  Yuan, Feng; Yu, Zhaolong; Ho, Luis C., 2009, \apj, 703, 1034

\bibitem[Yuan \& Narayan(2014)]{yn14} Yuan, F., \& Narayan, R. 2014, ARA\&A, 52, 529


\bibitem[Zhang(2013)]{zhang13} Zhang, S.-N.  2013, FrPhy, 8, 630                                       

\bibitem[Zdziarski et al.(2004)]{zd04} Zdziarski, A. A., Gierlinski, M., Mikolajewska, J., et al. 2004, \mnras, 351, 791

\bibitem[Zdziarski et al.(1996)]{zd96} Zdziarski, A. A., Johnson, W. N., \& Magdziarz, P. 1996, \mnras, 283, 193

\end{thebibliography}
\end{document}